\documentclass[10pt,a4paper]{article}
\usepackage{graphpap,epsfig,amssymb,graphicx,textcomp}
\begin{document}
\begin{center}
 \large{\bf SPATIO-TEMPORAL DYNAMICS OF THE KURAMOTO-SAKAGUCHI MODEL WITH TIME-DEPENDENT CONNECTIVITY}
\end{center}

\vskip 1cm

\begin{center}{\it Amitava Banerjee$^{1}$ and Muktish Acharyya$^{2}$}\\
\vskip 0.2 cm
{\it Department of Physics, Presidency University,}\\
{\it 86/1 College street, Kolkata-700073, INDIA}\\
\vskip 1 cm
{1. E-mail: amitava8196@gmail.com}\\
{2. E-mail: muktish.physics@presiuniv.ac.in}\end{center}

\begin{abstract}
We have studied the dynamics of the paradigmatic Kuramoto-Sakaguchi model of identical coupled phase oscillators with various kinds of time-dependent connectivity using Eulerian discretization. We first explore the parameter spaces for various types of collective states using the phase plots of the two statistical quantities, namely the strength of incoherence and the discontinity measure. In the quasi-static limit of the changing of coupling range, we have observed how the system relaxes from one state to another and have identified a few interesting collective dynamical states along the way. Under a sinusoidal change of the coupling range, the global order parameter characterizing the degree of synchronization in the system is shown to undergo a hysteresis with the coupling range. Finally, we study the low-dimensional spatio-temporal dynamics of the local order parameter in the continuum limit using the recently-developed Ott-Antonsen ansatz and justify some of our numerical results. In particular, we identify an intrinsic time-scale of the Kuramoto system and show that the simulations exhibit two distinct kinds of qualitative behavior in two cases when the time-scale associated with the switching of the coupling radius is very large compared to the intrinsic time-scale and when it is comparable with the intrinsic time-scale. 
\end{abstract}

\vskip 2cm


\twocolumn \noindent {\bf I. INTRODUCTION}
\vskip 0.5 cm

Examples of spontaneous synchronization of phases corresponding any rhythmic activity or oscillations of interacting individuals of a dynamical system abound in nature. Some notable instances from the living world showing phase-synchronous behavior are flashing of fireflies \cite{mat1,aria1}, chirping of crickets \cite{yeu1}, firing of pulse-coupled excitable neurons \cite{masuda,mirollo} and conduction of pacemaker cells \cite{mich} in the heart among many. Systems arising from elsewhere showing synchronous behavior vary in diversity and include laser arrays \cite{david1,vlad1}, microwave oscillators \cite{jung1,boris1} and superconducting Josephson junction arrays \cite{vlasov1,caw1} to mention a few. A paradigmatic model which is simple yet capable of capturing almost all the essence of synchronization of oscillations is the Kuramoto model which describes the temporal evolution of phases  of $N$ coupled oscillators by the following set of ordinary differential equations \cite{kuramoto1}

$$\dot{\theta_{i}}= \omega_{i}+\sum_{j=1}^{N}K_{ij}sin(\theta_{j}-\theta_{i}) \eqno(1.1)$$

where $\theta_{i}$ is the instantaneous phase of the $i-$th oscillator, $\omega_{i}$ being its natural frequency of oscillation selected randomly from a normalized probability distribution $g(\omega)$. The matrix $K_{ij}$ denoting the strength of coupling between pairs of oscillators captures the topology of connectivity. This model is now quite thoroughly studied \cite{strogatz4,acebron} and is seen to produce some characteristic collective dynamical states \cite{acebron} :- (i) the asynchronous state, where the phases evolve incoherently and which occurs for coupling strength lower than a critical value, (ii) the global synchronous state, where all the phases become almost equal after a certain time and evolve together as a mass thereafter, and the recently found \cite{kuramoto2,panaggio} (iii) chimera and multichimera states, which occur in identical oscillators for certain non-local coupling, where the oscillators separate into a coexistence of one or more synchronous groups, spaced by the rest of oscillators evolving asynchronously. This simple model showing such a rich spectrum of dynamical behavior has been successfully applied, sometimes slightly modified, to model a wide variety of natural and laboratory systems \cite{acebron} -- ranging from neuronal oscillations in the human cortex \cite{somp1,somp2,mich1}, Josephson Junction \cite{wies1,wies2}  and laser arrays \cite{olivia1}, charge density transport in metals \cite{str1}, coupled chemical oscillators \cite{kuramoto1}, spin glasses with random couplings \cite{isa1} and even earthquake sequences \cite{vasu1} to mention a few. 

     The Kuramoto model has been modified many times to accommodate for more general dynamics \cite{acebron}. Two of very popular modifications are inclusion of inertia in the model by adding a second order derivative term \cite{ace2,ace3,fila1,peng1,sham1,olmi1,acebron} and considering time-dependent coupling or forcing parameters in the model \cite{aoki1,anton1,cumin1,sch1,dav1,maes1,acebron}. The former  has proven to be quite useful in describing systems like power grids \cite{fila1}, disordered arrays of underdamped Josephson junctions \cite{stro1,acebron} etc. The latter, which is related to our present study, is useful in modeling many realistic effects. For example, bacteria or single-celled eukaryotes communicating through an external time-varying chemical medium \cite{dav2}, synaptic plasticity of neurons \cite{seli1}, circadian rhythm of plants and animals which respond to the periodically varying natural parameters or pedestrians walking on a common bridge \cite{stro2} are some of the systems coupled with time-varying external forcing. On the other hand, within the system itself, consideration of time-dependent coupling strength \cite{cumin1}, variable natural frequency of individual oscillators \cite{cumin1} or noisy couplings \cite{acebron,park1} constitute interesting problems. All of these being properties of realistic systems, are present in any natural system with varied amounts. There has been much development in these directions involving identification of different collective dynamical states and finding the conditions for their stability and bifurcations, using both numerical and analytical studies. A recent work \cite{spa1} derives and summarizes key results for the Kuramoto dynamics with time-dependent parameters. 
     
     However, very few of these studies involve a thorough analysis of the dynamics of the system having time-dependence on the coupling topology itself, i.e., a system, where the coupling radius or interaction range of individuals changes over time. Two such studies are indeed done \cite{lee1,so1}, however one is a brief report is essentially a numerical study and consider periodically switching couplings with random neighbors with the coupling radius varying as a triangular wave and also assumes a distribution of intrinsic frequency of oscillators \cite{lee1}. The other study concentrated the most on the synchronizing transition rather than spatio-temporal behavior of the system \cite{so1}. In this work, we have considered a much simple situation consisting of identical oscillators, each of which varies its range of connection in a continuous sinusoidal manner. Even this simple system is shown to reproduce the key results obtained in those earlier works \cite{lee1} and moreover, we are able to propose an analytical justification of these behaviors. 
     
     We propose some situations, where the consideration of time-dependent connectivity may become important. It is well-known that the synaptic connection pattern in the human nervous system changes over time and hence the number of synapses in the brain varies \cite{rak1,bou1} -- over the total lifespan as a normal process, during certain special circumstances and sometimes in older ages often due to aging-related neuro-degenerative diseases. We have already mentioned that Kuramoto model is extensively used to model neuronal activities, thus it will be interesting to know how this changes the large-scale firing behavior of a collection of neurons. The variable coupling radius may also arise in bacterial populations in which individuals communicate via some chemical signalling \cite{dav2} molecules whose concentrations may change over time. Also, in a somewhat different context, time-dependence in human connectivity arises in social and multimedia interactions and influences \cite{clau1} whenever there is a major event which occurs periodically (like elections, tournaments etc.) and triggers excitement, which slowly dies out after its completion. These motivate us to study the present problem.
     
     In this work, we consider identical oscillators with a varying coupling range for each. First we numerically simulate the spatio-temporal dynamics of the oscillators. We first study a quasi-steady variation of coupling radius, where the interaction radius is changed very slowly and each time coupling radius is changed, the system is allowed to come to an equilibrium before the next change. In this limit, we observe twisted states, whose multi-stability is described earlier in literature \cite{gir1}. We note that changes in coupling radius may or may not result in a change of their periodicity and often a hysteresis in their structural nature, which is visually evident, is involved. If we work in a parameter range so as to involve chimera and multichimera states in this situation, interesting dynamical states are seen to form, for example, a periodic multichimera. This also invites us to explore the dynamics with time-varying parameters and their effects to chimera states in other systems (like coupled Stuart-Landau or Van der Pol Oscillators) with a hope to discover yet new interesting collective states.
     
     Next, we change the coupling radius in a faster rate sinusoidally. The system periodically changes its degree of synchronization and synchronized clusters of various sizes form and break periodically. The global order parameter is seen to have a hysteretic behavior with the coupling radius and the area of the hysteresis loop is seen to increase with both the coupling strength and the peak value of the coupling radius.
     
     Lastly, we describe an analytic treatment for the spatio-temporal dynamics of the local order parameter in the thermodynamic ($N\rightarrow \infty$) limit using the recently developed Ott-Antonsen ansatz \cite{ott1}. This enables us to describe the dynamics in terms of only two coupled equations. Solving those numerically, we show that the system becomes homogeneous for low frequency variation of the coupling range or higher values of the coupling strength, but for higher frequencies, the system becomes inhomogeneous and the local order parameter varies from point to point. The hysteretic behavior of the order parameter is also reproduced.
       
     Before we proceed to the next section, let us clarify the difference between the present problem with those involving time-dependent coupling strengths studied in detail earlier. The present study matches with them in the mean-field limit, where we neglect individual contributions to the dynamics and may accommodate the variable coupling radius by varying the mean-field interaction strength accordingly. However, this approach being a mean-field one, fails to consider spatial inhomogeneity in the system, which will be important for certain parameter values in our case, as we shall see. So, a finite range of coupling will not be a faithful sample of the total system and moreover, total interactions will vary considerably among different individuals. Thus mean-field approximation will not be a very good one in general.
    
    This paper is organized as follows :- Section II describes various dynamical states found in the equilibrium condition, i.e., with a constant coupling radius and distinguishes between them using some statistical measures of discontinuity and strength of incoherence. Section IIIa describes the behavior of these states under a quasi-static change in coupling and introduces some interesting dynamical states and the hysteretic transformation among the variety of states. The next part of the section (IIIb) explores the nonequilibrium phase dynamics and the time-variable distribution of the synchronous cluster sizes for a sinusoidally varying coupling radius and its effects on the global order parameter. The following section (IV) uses the recent Ott-Antonsen ansatz to identify the low-dimensional spatio-temporal dynamics involving the local order parameter and provides some analytic support for the observed behavior in section IV. Finally, the concluding section summarizes the result and discusses about some future directions and possible extensions of the present work. 

\vskip 1 cm

\noindent {\bf II. PHASE DIAGRAM FOR THE EQUILIBRIUM LIMIT}
\vskip 0.5 cm
In this paper, we study a slightly general version of equation $(1.1)$, namely
the Kuramoto-Sakaguchi model, described by the following coupled differential
equation

\begin{equation}
{{d \theta_i} \over dt} = \omega - 
{{\epsilon} \over {2R}}
\sum_{j=i-R}^{j=i+R} {\rm sin}(\theta_i(t)-\theta_j(t)+\alpha)
\end{equation}
\noindent where, $\theta_i(t)$ is the phase of i-th oscillator at time $t$.
$\omega$ is the intrinsic frequency of each oscillator which could be set equal
to zero by a rotating frame transformation without loss of generality. $\epsilon$ is the strength of coupling. 
$R$ is the radius of coupling, which means, 
phase of the i-th oscillator is coupled to $R$ number of oscillators in each
direction and $\alpha$ is Sakaguchi phase factor. The radius of coupling may be defined as $r=\frac{R}{N}$, where $N$ is the total number of oscillators. The boundary 
condition applied here is periodic. 

We solve this differential equation (with $\omega=0$) for $100$ oscillators by Eulerian 
discretization over time with $dt=1.0$. The forms of the differential equations thus
become

\begin{equation}
\theta_i(t+1) = \theta_i(t) - 
{{\epsilon} \over {2R}}
\sum_{j=i-R}^{j=i+R} {\rm sin}(\theta_i(t)-\theta_j(t)+\alpha).
\end{equation}

\noindent Starting from a random initial values of
the phases ($\theta_i$s) uniformly distributed between angles $0$ to $2\pi$, we have solved the above set of equations by simple numerical iteration method. For various set of the parameters $\epsilon$, $R$ and $\alpha$ we observe the usual dynamical states -- asynchronous, synchronous, chimera and multichimera. To have a complete idea about various dynamical states in the total parameter range, we use two statistical measures defined in earlier works \cite{gopal1,gopal2} to characterize them.  The first one of them is the
parameter $S$, the strength of incoherence, which can be calculated according to the following prescription\cite{gopal1,gopal2}: at a particular time step, we can define a new variable $Z_i=P_i-P_{i+1}$ to work with, which is nothing but the difference of the phase of the $i-$th oscillator from its nearest neighbor to the right. Then one divides the total of $N$ oscillators into $N_g$ groups side-by-side, so that
each group contains $N_b=N/N_g$ number of oscillators. Next one calculates, the time-averaged
variance of $Z_i$ in k-th group by the equation:

\begin{equation}
\sigma_k= \Big\langle \sqrt{{\sum_{j=1}^{N_b}(Z_i^k-{\bar Z_i^k})^2} \over {N_b}}\Big\rangle_{t}.
\end{equation}

\noindent Next a new variable for each group is defined as
\begin{equation}
\lambda_k=\Theta (\delta - \sigma_k)
\end{equation}

\noindent where $\delta$ is a sufficiently small value and $\Theta$ represents the Heavyside step
function. Finally  with the help of this, the strength of incoherence $S$ may be defined as

\begin{equation}
S = 1 - {{\sum_{k=1}^{k=N_g} \lambda_k} \over {N_g}}.
\end{equation}

\noindent Here, $S=1$ represents complete incoherence, $S=0$ represents
complete coherence and $0<S<1$ represents chimera or multichimera states
of the coupled oscillators. 
\begin{figure}[h]
\begin{center}
\begin{tabular}{c}
        \resizebox{6cm}{!}{\includegraphics[angle=-90]{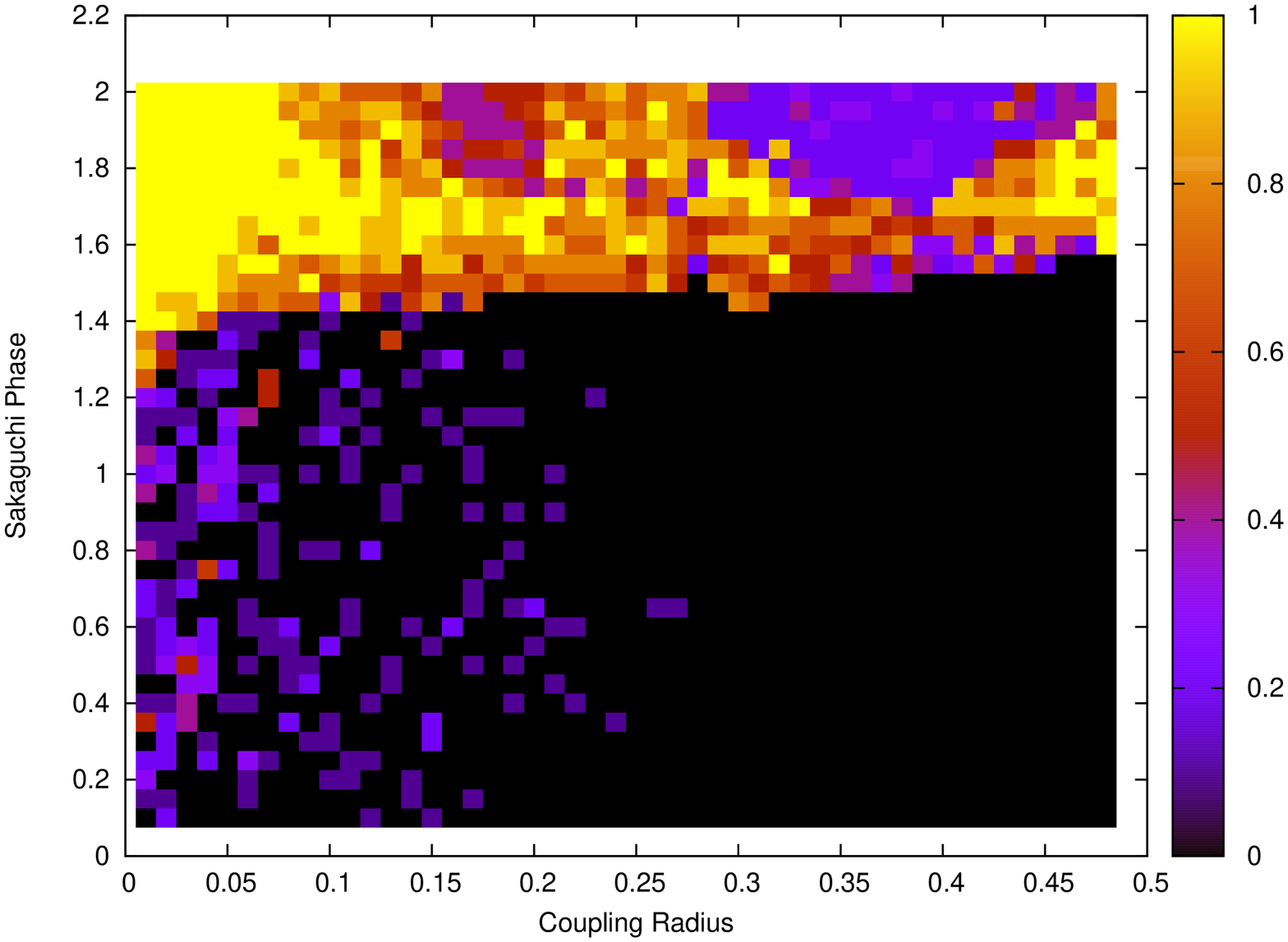}}\\
        \resizebox{6cm}{!}{\includegraphics[angle=-90]{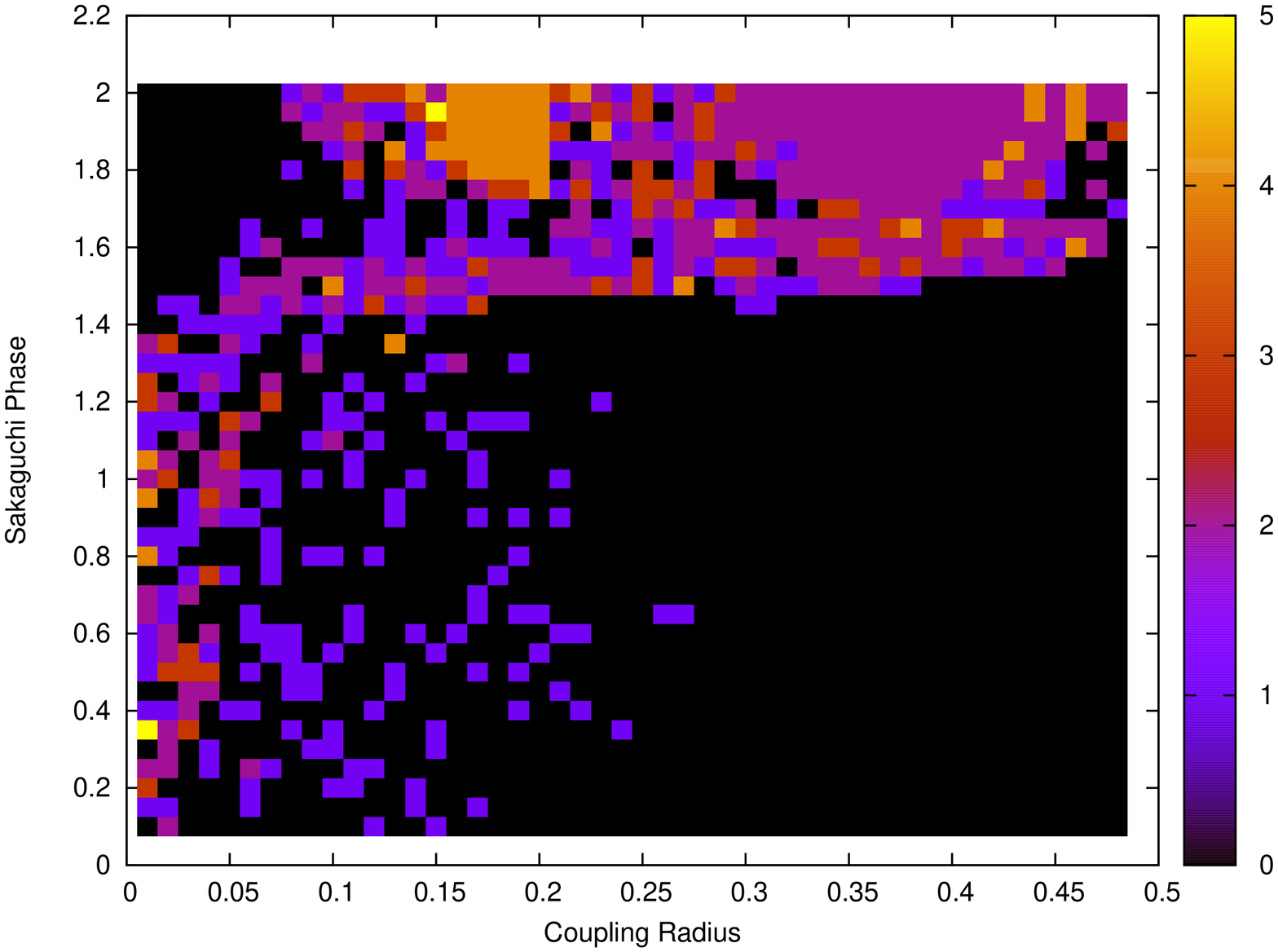}}
                  \end{tabular}
 \caption{\tiny{Phase Plot for Strength of Incoherence (top) and Discontinuity Measure (bottom) for $\epsilon=0.5,N=100$ after $5000$ iterations and $\delta=1.0$, note the sharp change near $\alpha=\frac{\pi}{2}$.}}
\end{center}
\end{figure}
To distinguish further between single chimera and multichimera states
(as for both the cases $0<S<1$) one resorts to finding the discontinuity measure defined as\cite{gopal1,gopal2}

\begin{equation}
\eta = \frac{(\sum_{k=1}^{k=N_g} |\lambda_k - \lambda_{k+1}|)}{2}
\end{equation}

\noindent with the periodic boundary identification $\lambda_{N_{g}+1}=\lambda_{1}$. 
Here, $\eta=0$ for both completely synchronous and completely asynchronous states, whereas $\eta=1$ for single chimera state and some positive integer
between $1$ and $\frac{N_{g}}{2}$ for multichimera states. 
 Thus it can be used for a more detailed characterization of chimera states. We plot the phase diagram for both of these measures in the $\alpha-r$ (fig. $1$) and $\epsilon-r$ (fig. $2$) planes (these are obtained for $100$ oscillators after $5000$ iterations, a sufficiently large time for the phase dynamics to set in a steady state so that these measures do not change much over time). Besides determining the parameter ranges for various dynamical states to occur, these plots reveal some interesting features about the dynamics. Firstly, we observe the important role of Sakaguchi phase factor ($\alpha$) in
the synchronization. A careful observation of fig. $1$ around $\alpha=\pi/2$ shows that for $\alpha > \pi/2$, the oscillators do not show any globally synchronized state, even for global couplings. So, to have global
synchronization, one has to choose $\alpha < \pi/2$ and in this work, we restrict ourselves to work mostly in this limit. This transition resulting in the loss of global synchrony near $\alpha=\pi/2$ is sudden and marked by the sharp boundary in the phase plots. As a result the $\epsilon-r$ plane plots are quite different qualitatively in these two regimes. 
\begin{figure}[h]
\begin{center}
\begin{tabular}{c}
        \resizebox{4.5cm}{!}{\includegraphics[angle=-90]{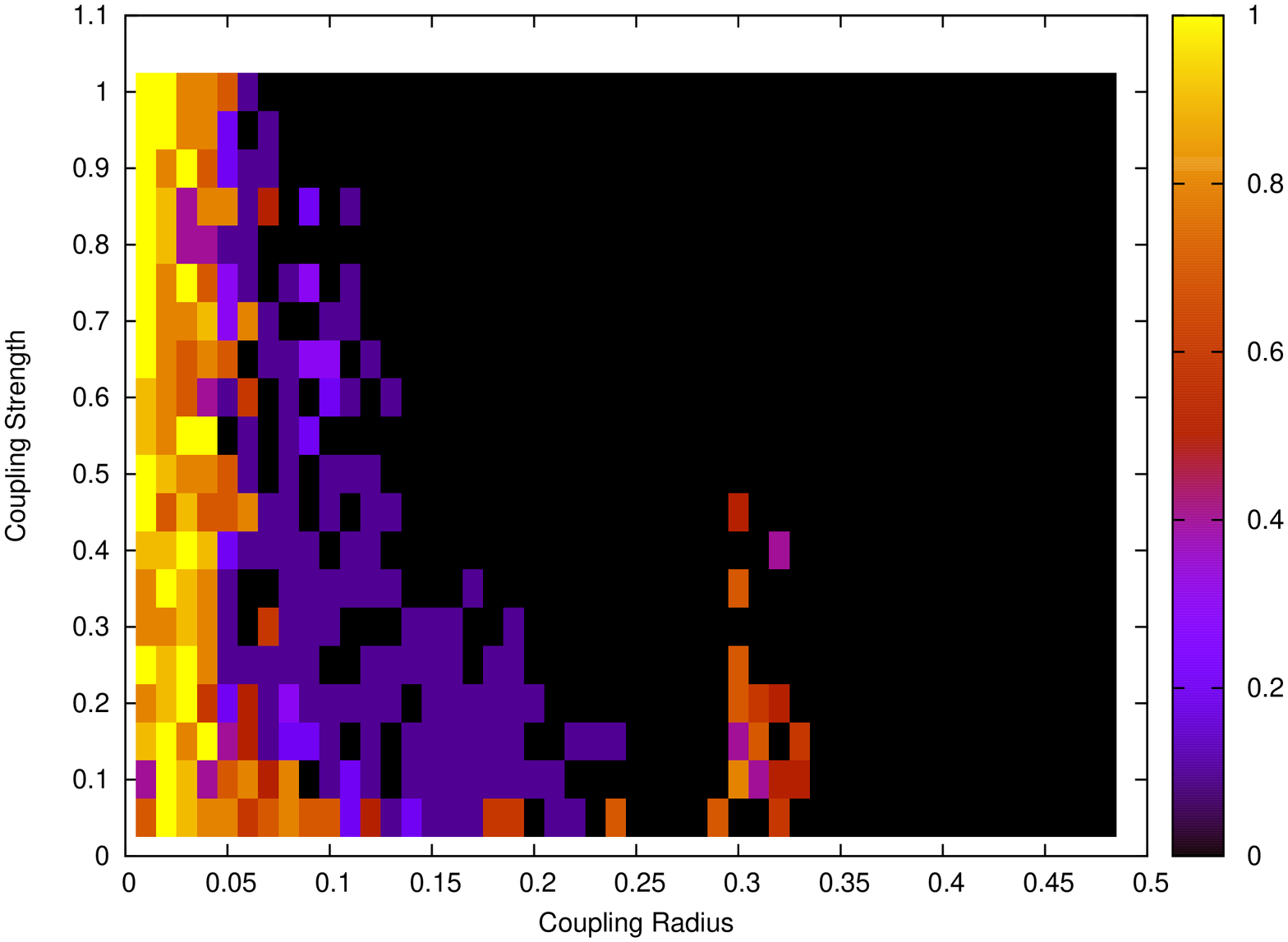}}
        \resizebox{4.5cm}{!}{\includegraphics[angle=-90]{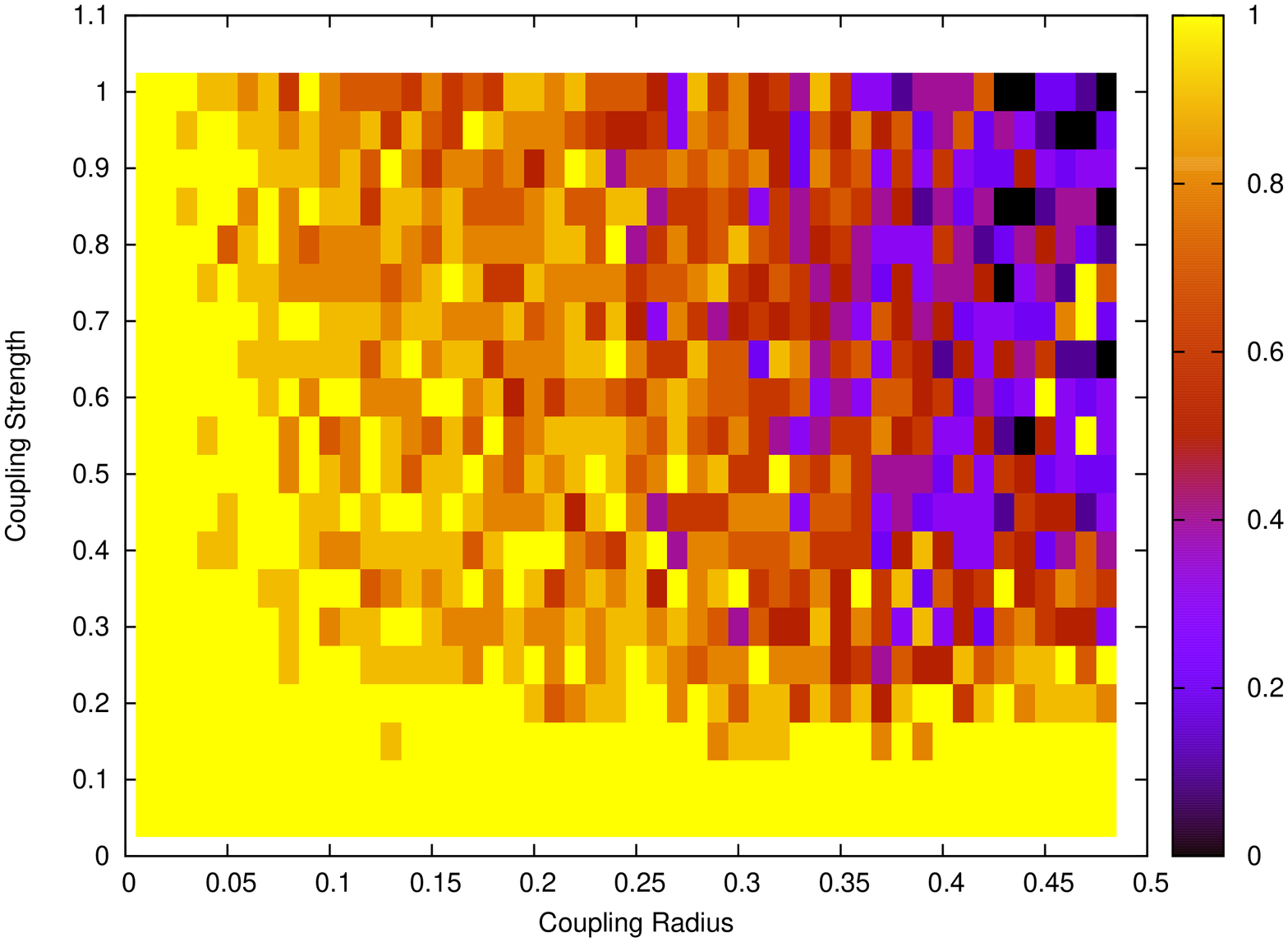}}
        \\
        \resizebox{4.5cm}{!}{\includegraphics[angle=-90]{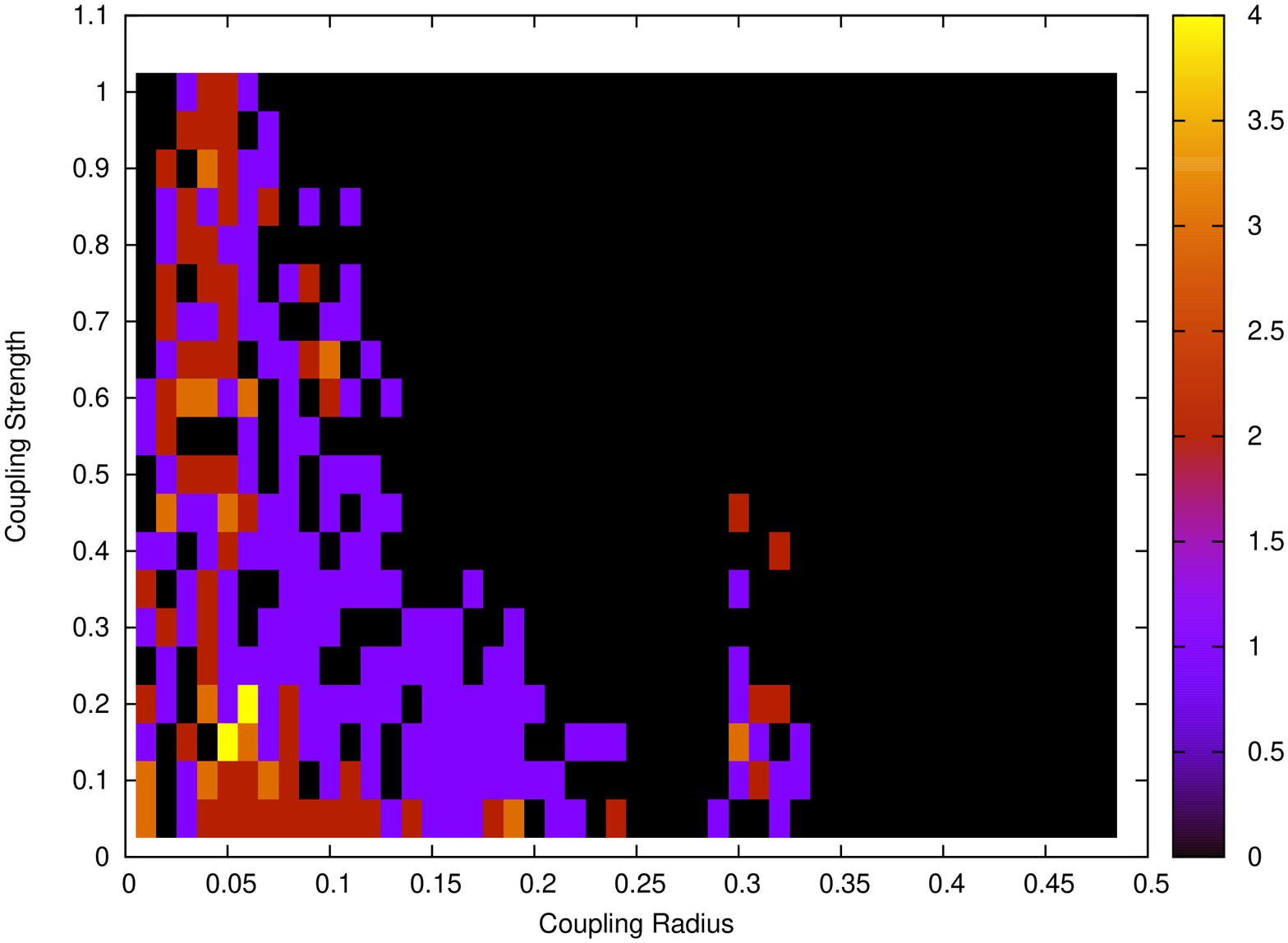}}
        \resizebox{4.5cm}{!}{\includegraphics[angle=-90]{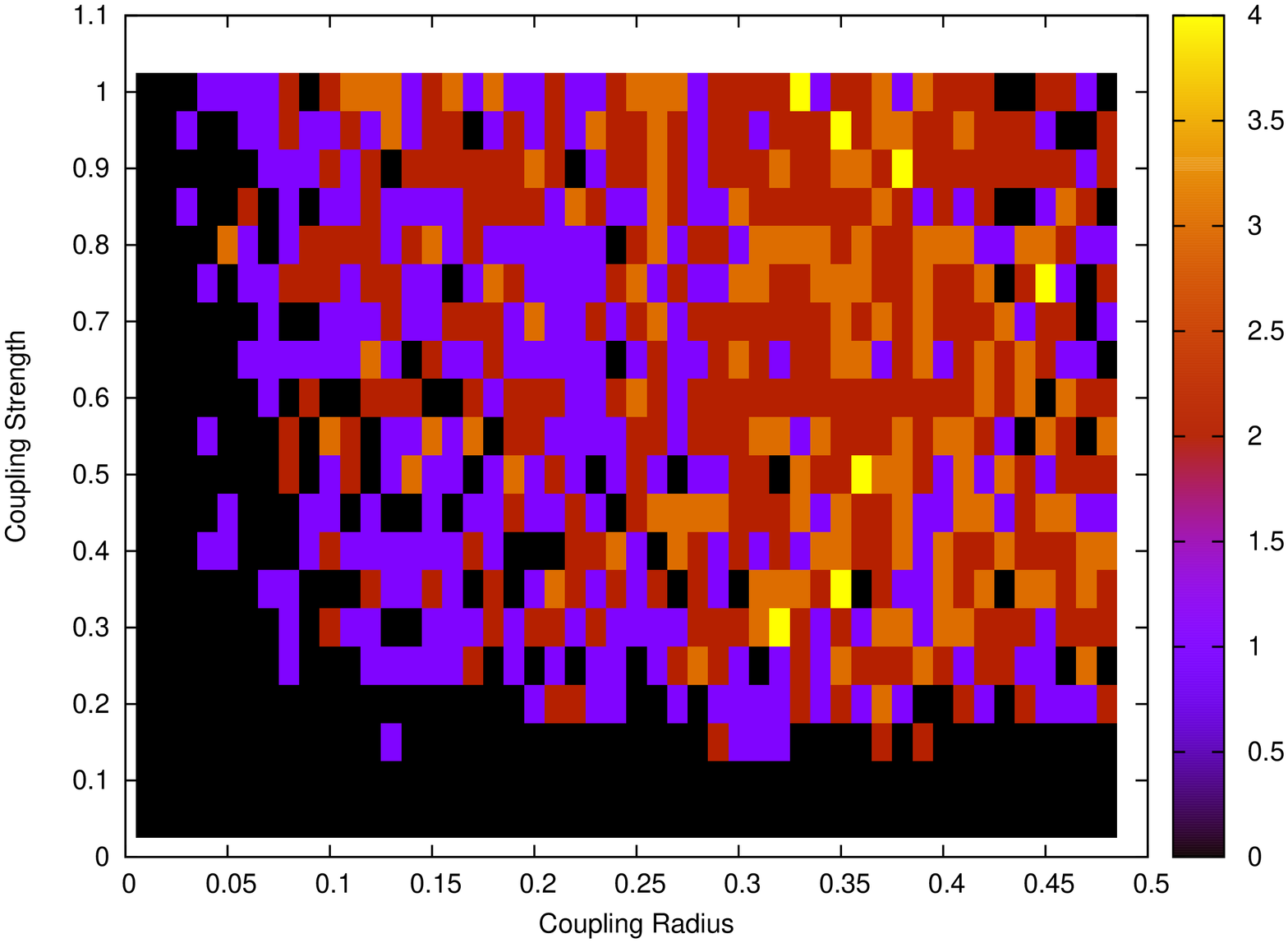}}
                  \end{tabular}
 \caption{\tiny{Phase Plots for Strength of Incoherence (top right and left) and Discontinuity Measure (bottom right and left) for $N=100$ after $5000$ iterations, $\delta=1.0$ with $\alpha=1.4$ (top and bottom left) and $\alpha=1.6$ (top and bottom right) note the qualitative changes across $\alpha=\frac{\pi}{2}$.}}
\end{center}
\end{figure}
Apart from the usual four dynamical states described above, the system shows various `twisted' or ordered states for certain parameter ranges (e.g., for $\epsilon=1.0,\alpha=1.0, r=0.46$), observed and analyzed previously in earlier works \cite{gir1}, where the phase difference of an oscillator with the next one is equal for each of them with the numerical value of this difference depending on the exact parameter values or the phase differences follow a simple regular pattern. All these various kinds of states will be involved in the discussion next section. We conclude this section with the list of various collective states obtained from random initial conditions shown in fig. $3$. 
\begin{figure}[h]
\begin{center}
\begin{tabular}{c}
        \resizebox{4cm}{!}{\includegraphics[angle=-90]{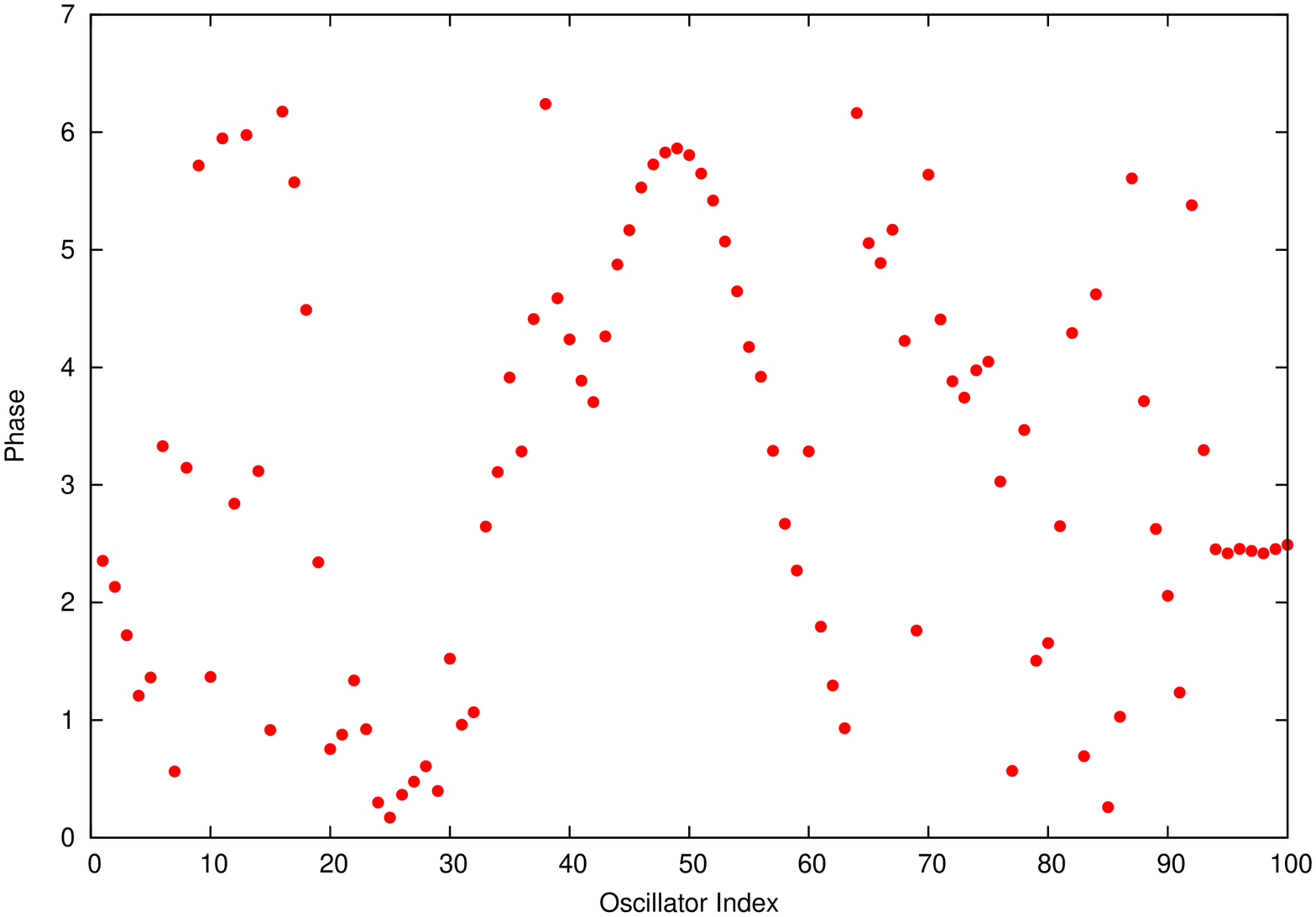}}
        \resizebox{4cm}{!}{\includegraphics[angle=-90]{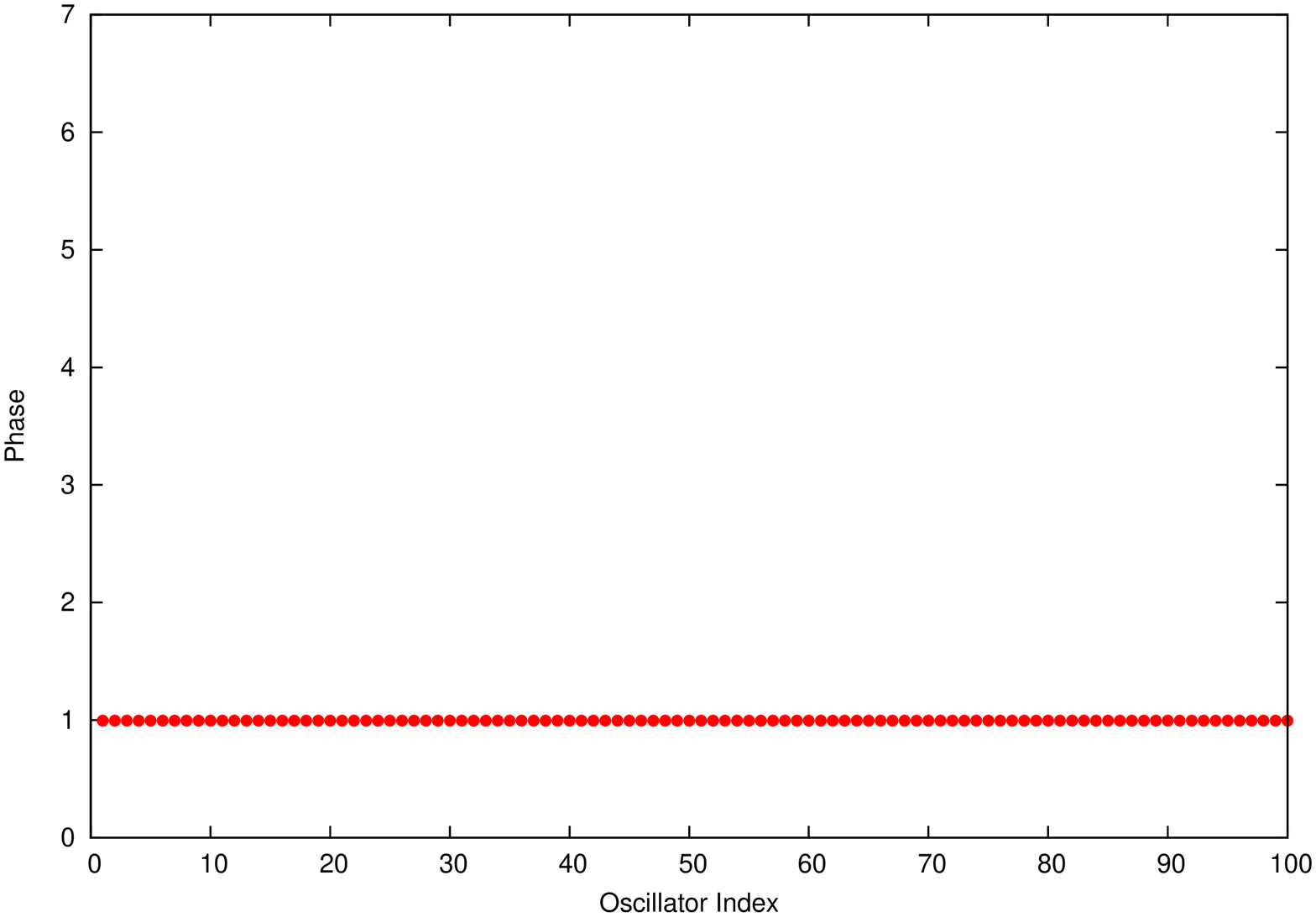}}
        \\
        \resizebox{4cm}{!}{\includegraphics[angle=-90]{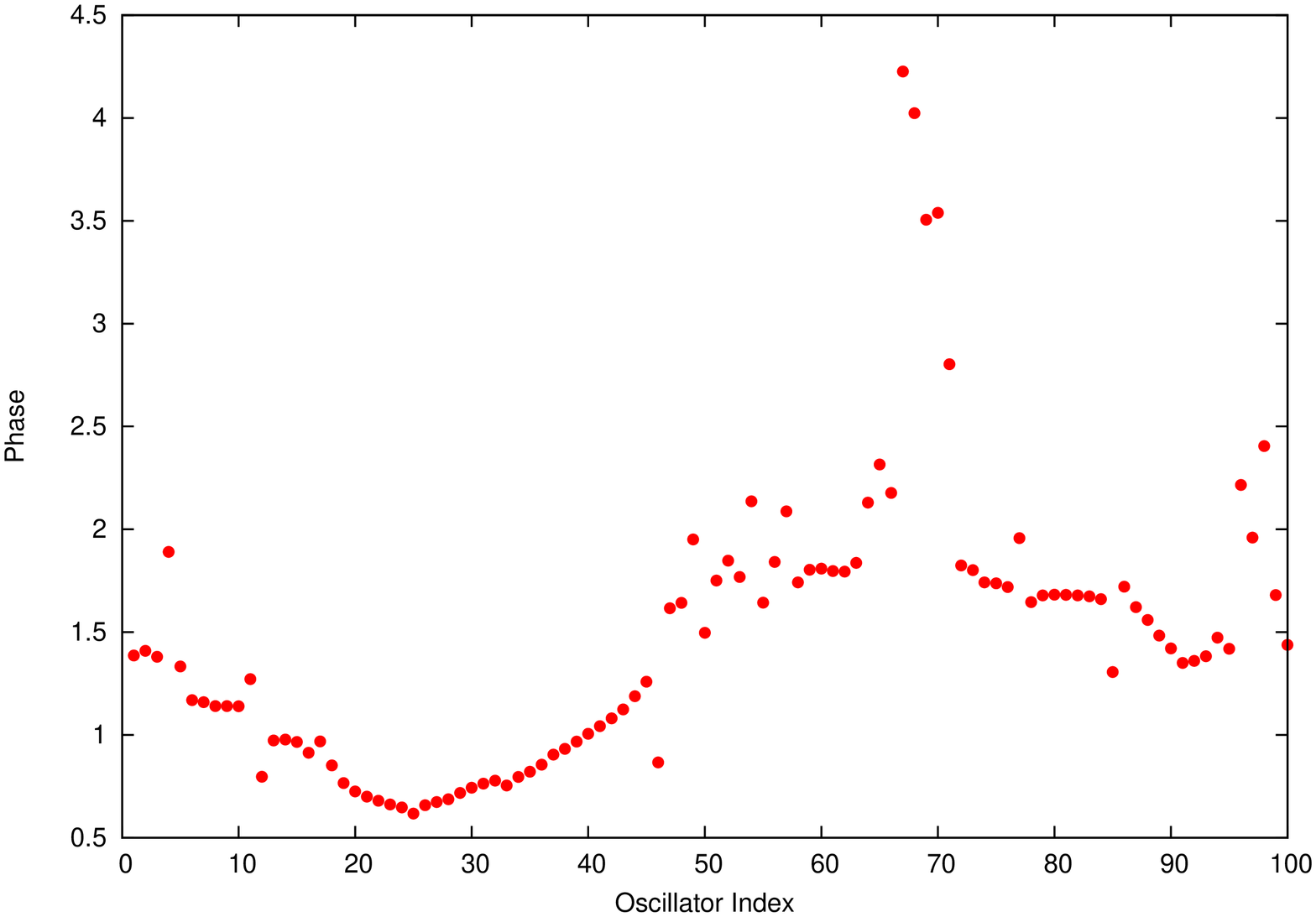}}
        \resizebox{4cm}{!}{\includegraphics[angle=-90]{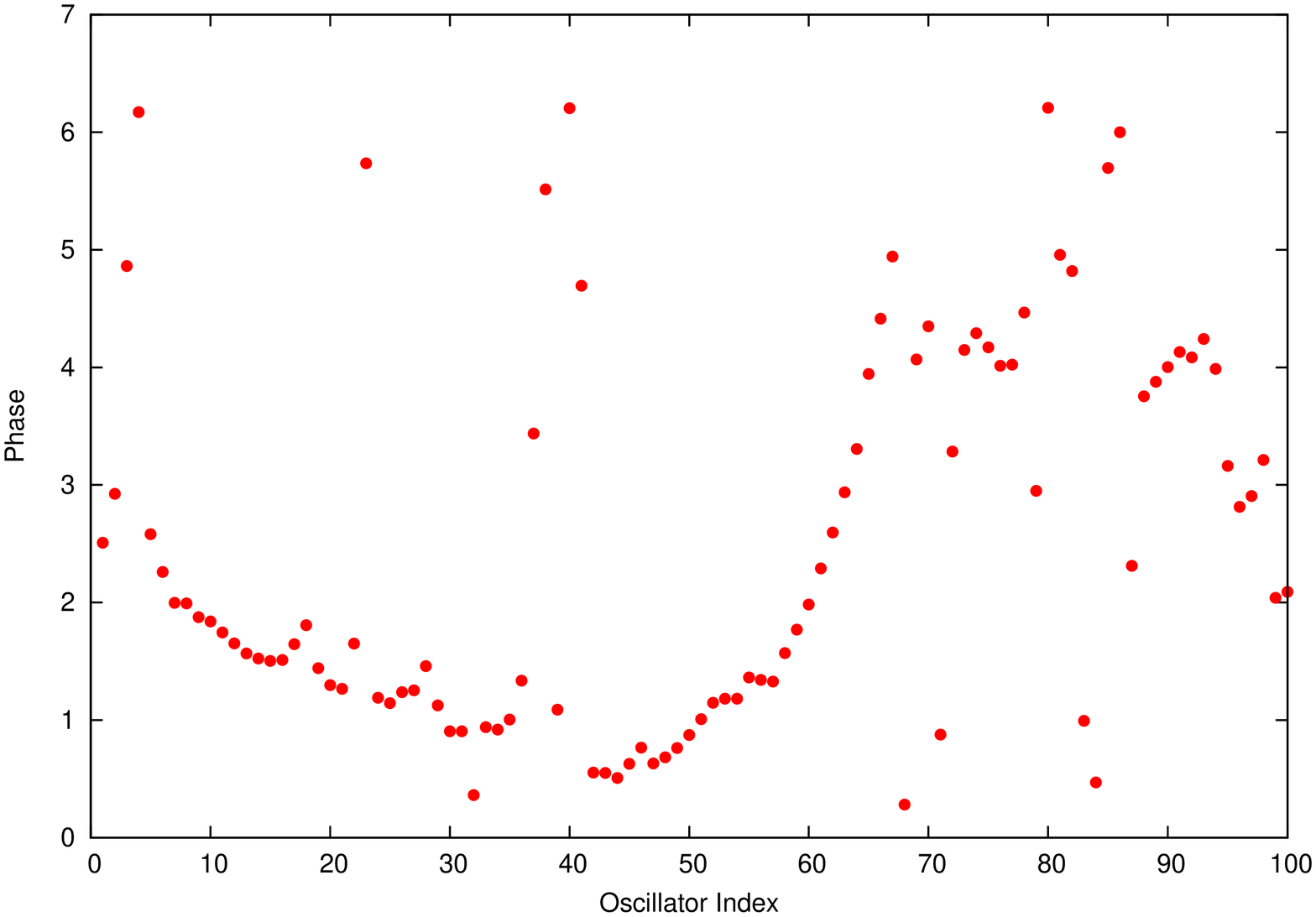}}
         \\
        \resizebox{4cm}{!}{\includegraphics[angle=-90]{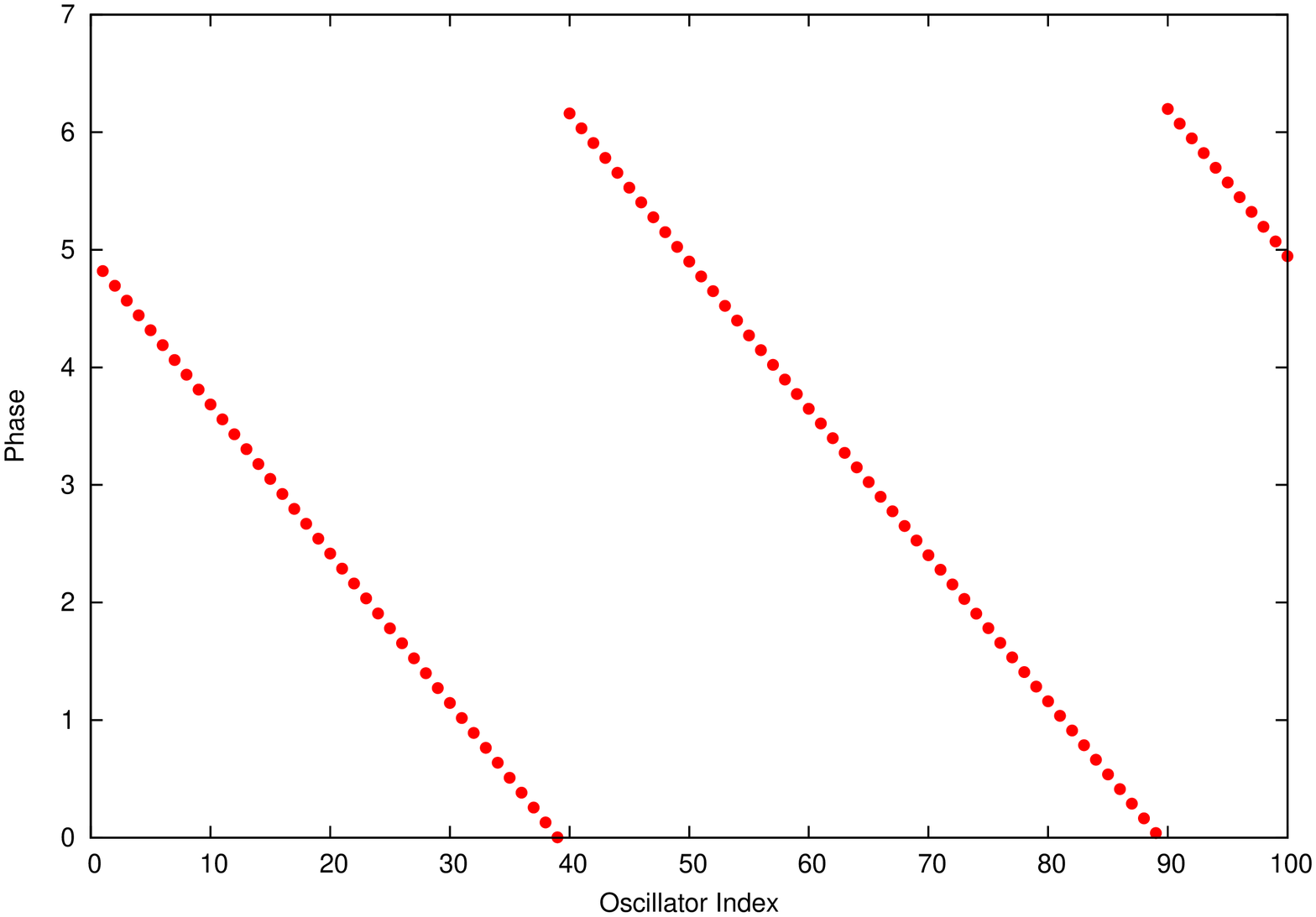}}
        \resizebox{4cm}{!}{\includegraphics[angle=-90]{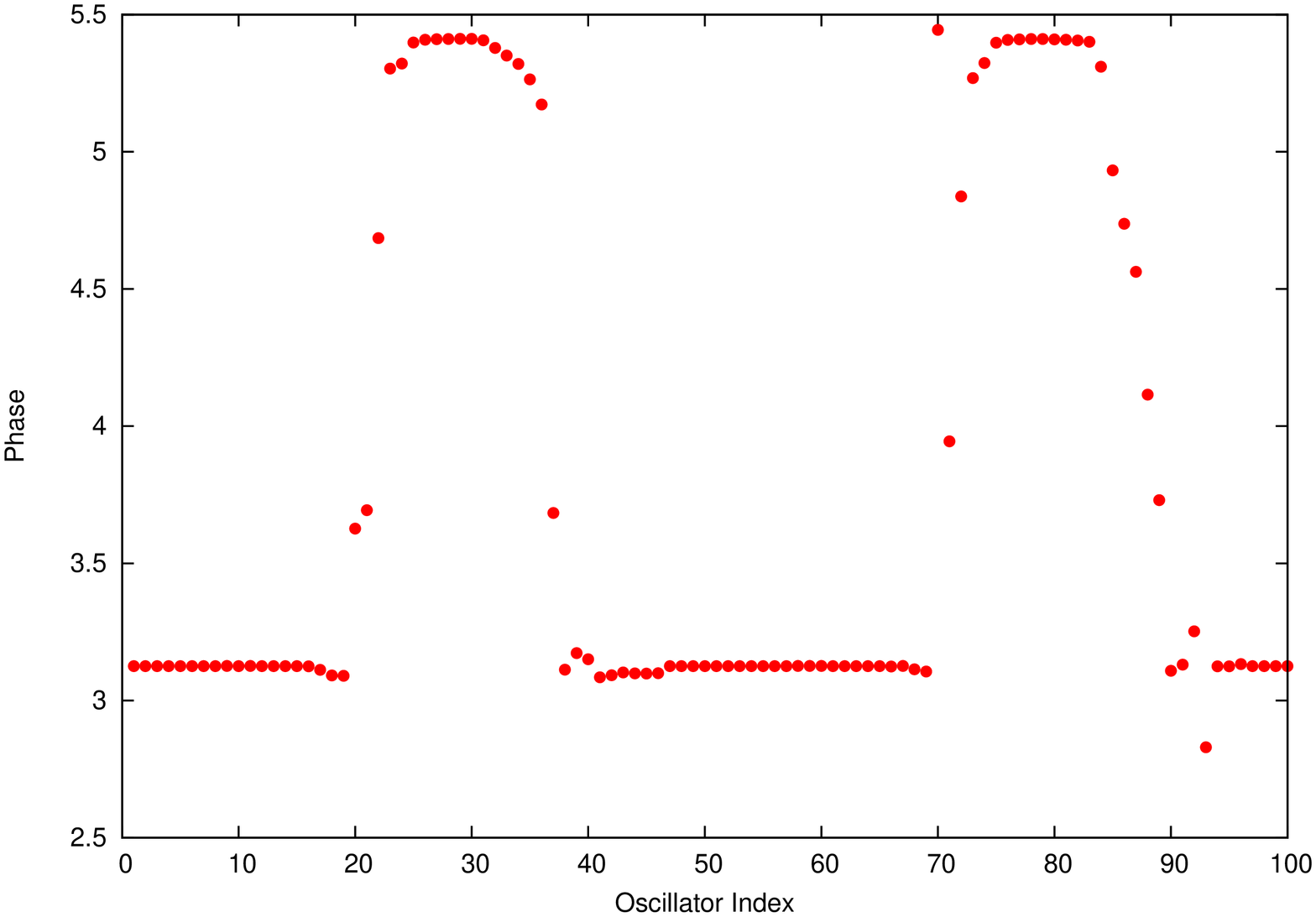}}
                  \end{tabular}
 \caption{\tiny{The Zoo of Collective States: Clockwise from top left -- Asynchronous ($\alpha=1.4,\epsilon=0.01,r=0.01$), Synchronous ($\alpha=1.4,\epsilon=1.0,r=0.49$), Multichimera ($\alpha=1.46,\epsilon=0.5,r=0.15$), Ordered ($\alpha=1.6,\epsilon=1.0,r=0.46$), Twisted ($\alpha=1.0,\epsilon=1.0,r=0.02$) and Chimera ($\alpha=1.46,\epsilon=0.9,r=0.2$) states all self-organized from a random initial state.}}
\end{center}
\end{figure}

\vskip 3.5 cm
 
\noindent {\bf III. DYNAMICS WITH VARIABLE COUPLING -- COLLECTIVE STATES AND HYSTERESIS}
\vskip 1 cm
\noindent {\bf (a) Dynamics with Two Values of Coupling Ranges}
\vskip 0.5 cm

In this section, we discuss the major features of the phase dynamics of the oscillators found by iterating eqn. $(2)$ over time with changing coupling radius, with a parameter regime where global synchronization do not take place. To this end, we first discuss about the quasi-static limit. In this regime, we keep the coupling radius fixed for a certain time which is sufficiently long (say, $5000$ iterations for $N=100$) such that the system comes to an equilibrium (characterized by little or no major qualitative change of the relative position of the phases over time) and then switch the coupling radius $r$ or the Sakaguchi phase $\alpha $ to another value instantaneously and see how the system responds. There it becomes evident that certain states (mainly some of the twisted states) are very stable under changes in $r$ and some are not. Moreover, during the relaxation after a switch, many of the states tend to retain some of the order of its past state(s) and so we propose that this method can be used to search for new self-organized collective dynamical states. For example, we inform about two interesting states -- one being a multichimera but with an exact periodic repetition (a ``crystalline chimera'') and another one a state with an well-defined short-range order in space (a ``liquid-like'' structure) shown in fig. $(4)$. A systematic study to find out the exact list of conditions leading to various dynamical states is yet to be done.
\begin{figure}[h]
\begin{center}
        \resizebox{3.5cm}{!}{\includegraphics[angle=-90]{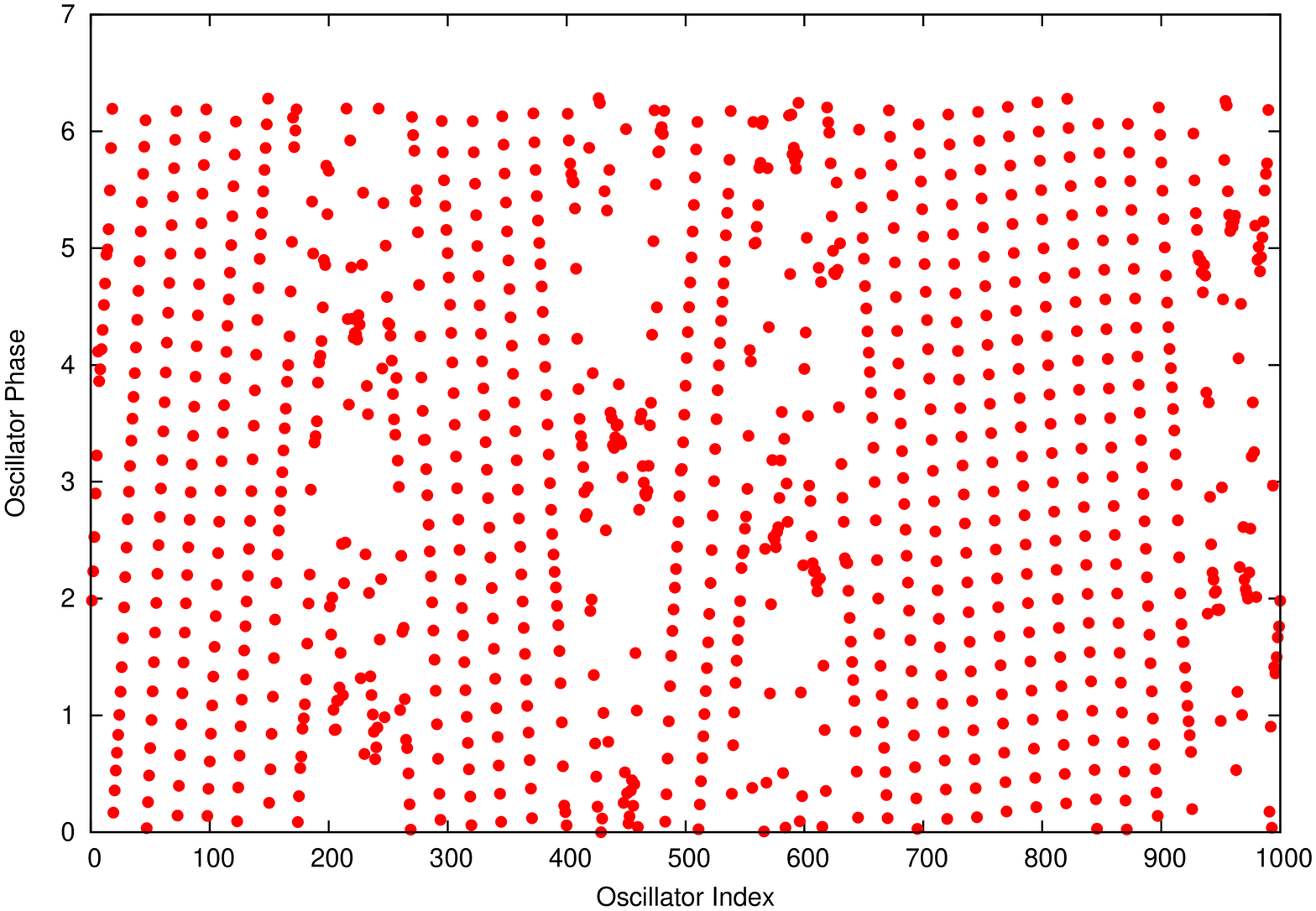}} 
        \resizebox{3.5cm}{!}{\includegraphics[angle=-90]{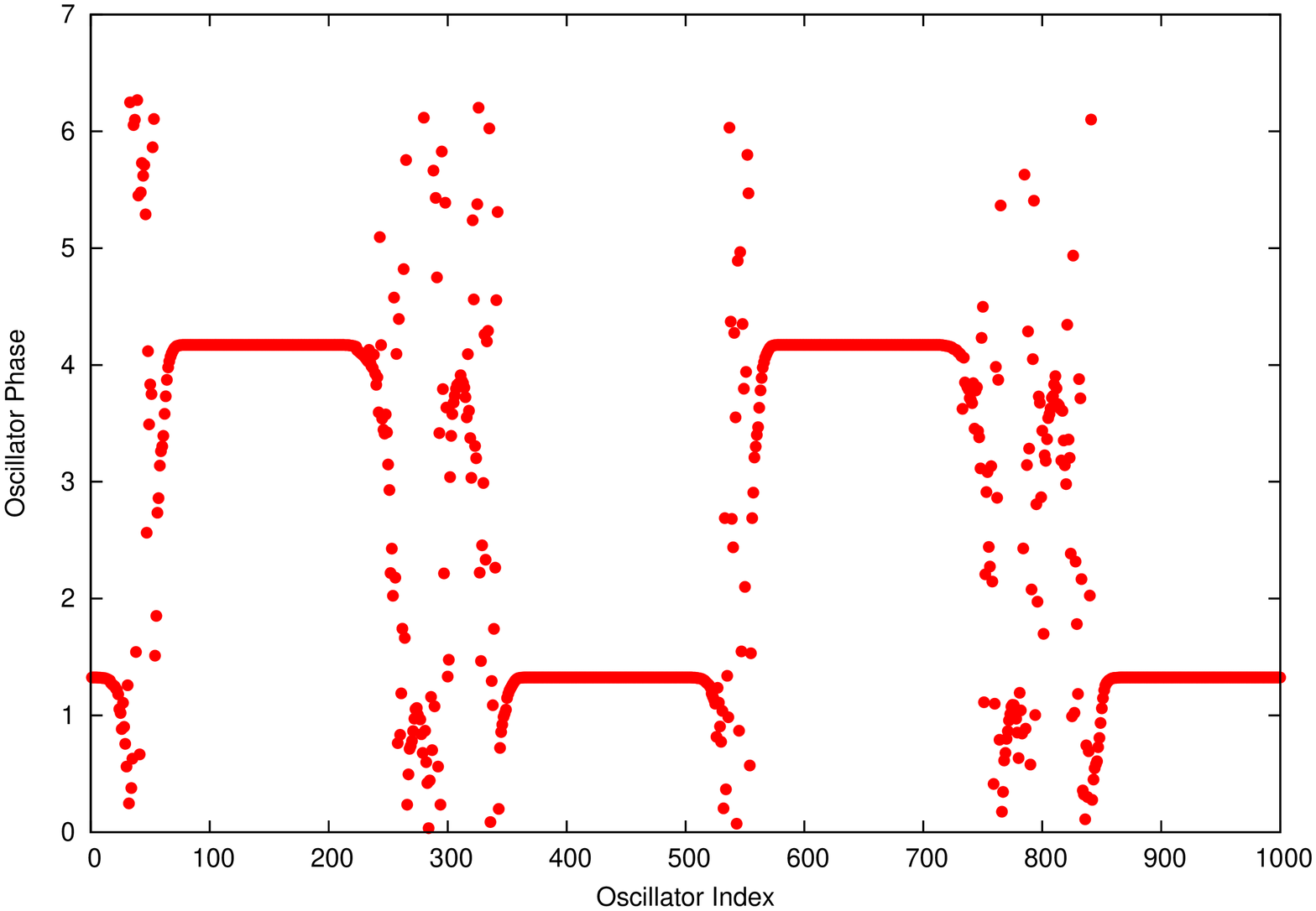}}

 \caption{\tiny{Collective States: Short Range Ordered State created from the random initial condition after sequentially using $r=0.12$ and $r=0.02$ each for $5000$ steps of iteration with $\alpha =\pi +1.0$ (left) and Periodic Multichimera created from the random initial condition after sequentially using $r=0.33$ for $20000$ steps and $r=0.49$ for $30000$ steps of iteration with $\alpha =\frac{\pi }{2}+0.11$ (right) both for $N=1000$. Note that none of these states are obtainable directly from random initial conditions with the set of parameter values used in the second step throughout.}}
\end{center}
\end{figure}
The states with a short-range order are quite interesting. To create them, starting from a random initial condition, we first form twisted states (these are the equilibrium states for a wide range of coupling radius values with $\alpha =\pi +1.0$, say) and then switch the coupling range to a lower value. There we note that the twisted states initially survive for some time. But perturbations over it are seen to grow and at some point of time it looks as if the phases are randomly distributed in the whole range. But then, the system reorders itself and the equilibrium state is found to have a short-range order. It consists of multiple domains of twisted states distinguished by varied nearest-neighbor phase differences separated by some randomly distributed phases at the interface as shown in the left side fig. $(5)$. Interestingly, the exact number of the domains of twisted states depend on the initial value of the coupling range determining the magnitude of the nearest-neighbor phase difference in the twisted states. To quantify the short-range spatial correlation, we use the autocorrelation function $A(i),i=1,2,...,\frac{N}{2}$ defined by the following equation
\vskip 0.5 cm
\begin{equation}
 A(p)=\left < (\theta (i)-\left< \theta \right>)(\theta (i+p-1)-\left< \theta \right>)\right>
\end{equation}
\vskip 0.5 cm
where $\theta(i)$'s are the phases of oscillators and $\left< \theta \right>$ is their average value. This function measures the spatial correlation of the phases for oscillators situated at distances $p$ apart and is plotted for various states in the right side of fig. $(5)$. In twisted state, the phases are totally ordered and so correlated in space, so the autocorrelation function does not decay with distance. For the case of multiple domains, the long-range correlation is generally lost and indicated by a decaying correlation profile. However, if there is a sufficiently large number of domains, then the phases in one domain may be correlated to those in another domain situated at some distance from the first. In that case, the autocorrelation profile would show a decay in the immediate vicinity of an oscillator, but a somewhat steady value at sufficiently larger distances. In general, the fact that different sorts of spatial correlations are found in the system for same value of final coupling radius demonstrates that the details of initial twisted states are retained somehow in the system and the latter's further phase dynamics is strongly affected by it. This is one of the important results involving time-dependent coupling.

\begin{figure}[h]
\begin{center}
\begin{tabular}{c}
        \resizebox{8cm}{!}{\includegraphics[angle=0]{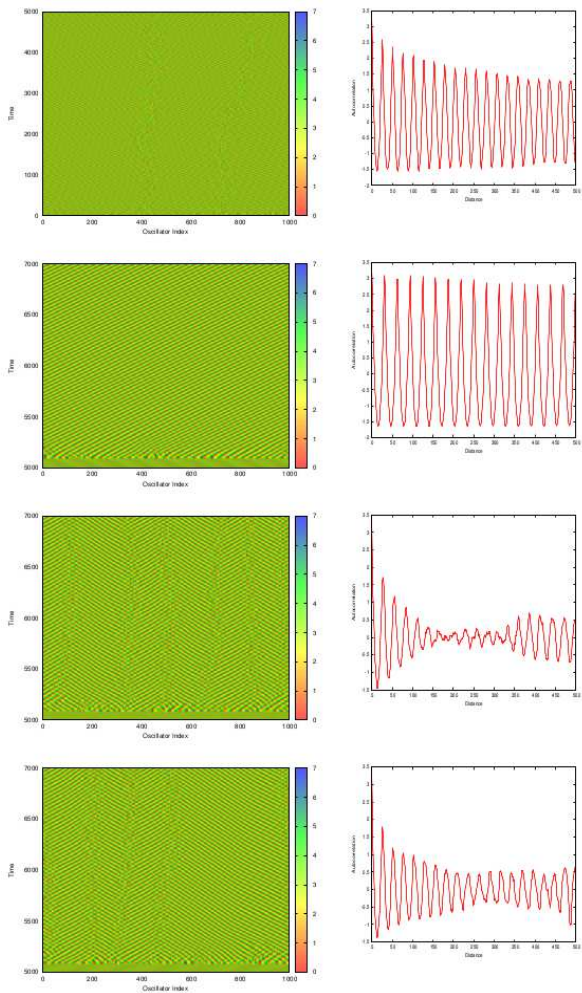}}

                  \end{tabular}
 \caption{\tiny{Phase Dynamics of Collective States with Short-Range Order and their Spatial Corrrelation Profile at the last time step: $L=1000,\alpha =\pi +1.0$ for all, the Coupling Radius ($r$) is switched to $0.02$ at $5000th$ step. The Coupling Radius at the First Step was (from top to bottom) $0.02,0.08,0.10$ and $0.12$. Note the initial Twisted States in all cases except the First one.}}
\end{center}
\end{figure}

\vskip 1 cm
\noindent {\bf (b) Dynamics with Continuously Varying Coupling Ranges}
\vskip 0.5 cm

Next we change the radius of coupling sinusoidally with time with an angular frequency $\omega _{r}$ (corresponding to a time-period $T_{r}$), i.e., take $R(t) = R_{0}+\lfloor R_{0} {\rm sin}(\omega_{r} t)\rfloor$ where the floor function $f(x)=\lfloor x \rfloor$ denotes the integral part of $x$. It becomes clear that due to the time dependent
coupling, the system oscillates between staes having various degrees of synchronization (fig. $(6)$ left). For a quantitative characterization, we calculate the time dependent global order parameter 
$Z(t)= {1 \over N} \sum_{k=1}^N e^{i\theta_k(t)}.$ The global order parameter $Z(t)$ has
a variation in time maintaining a phase difference with $R(t)$ (fig. $(6)$ right). 
\begin{figure}[h]
\begin{center}
\begin{tabular}{c}
        \resizebox{4cm}{!}{\includegraphics[angle=0]{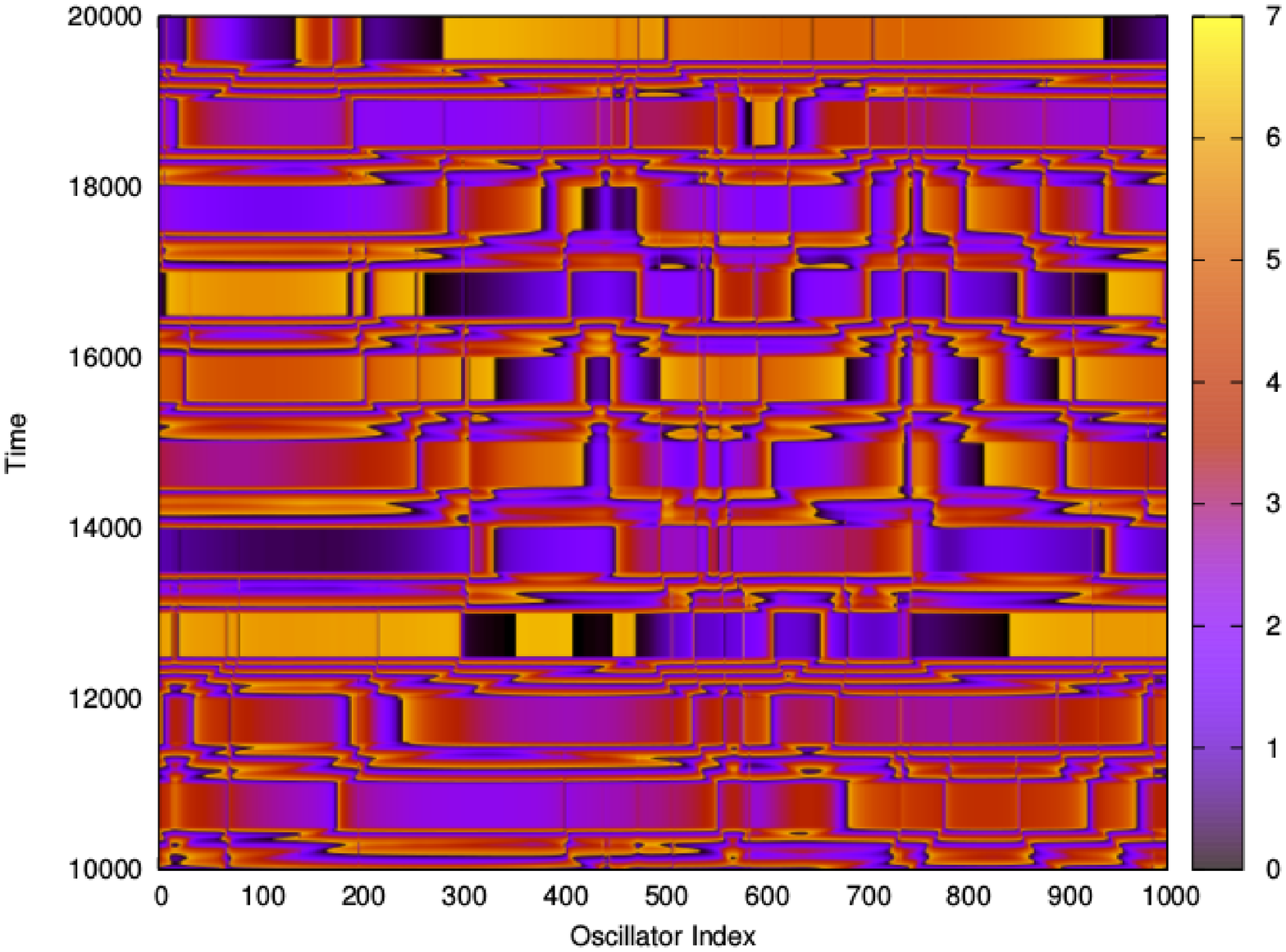}}\\
        \resizebox{!}{!}{\includegraphics[width=4cm,height=5cm, keepaspectratio,angle=-90]{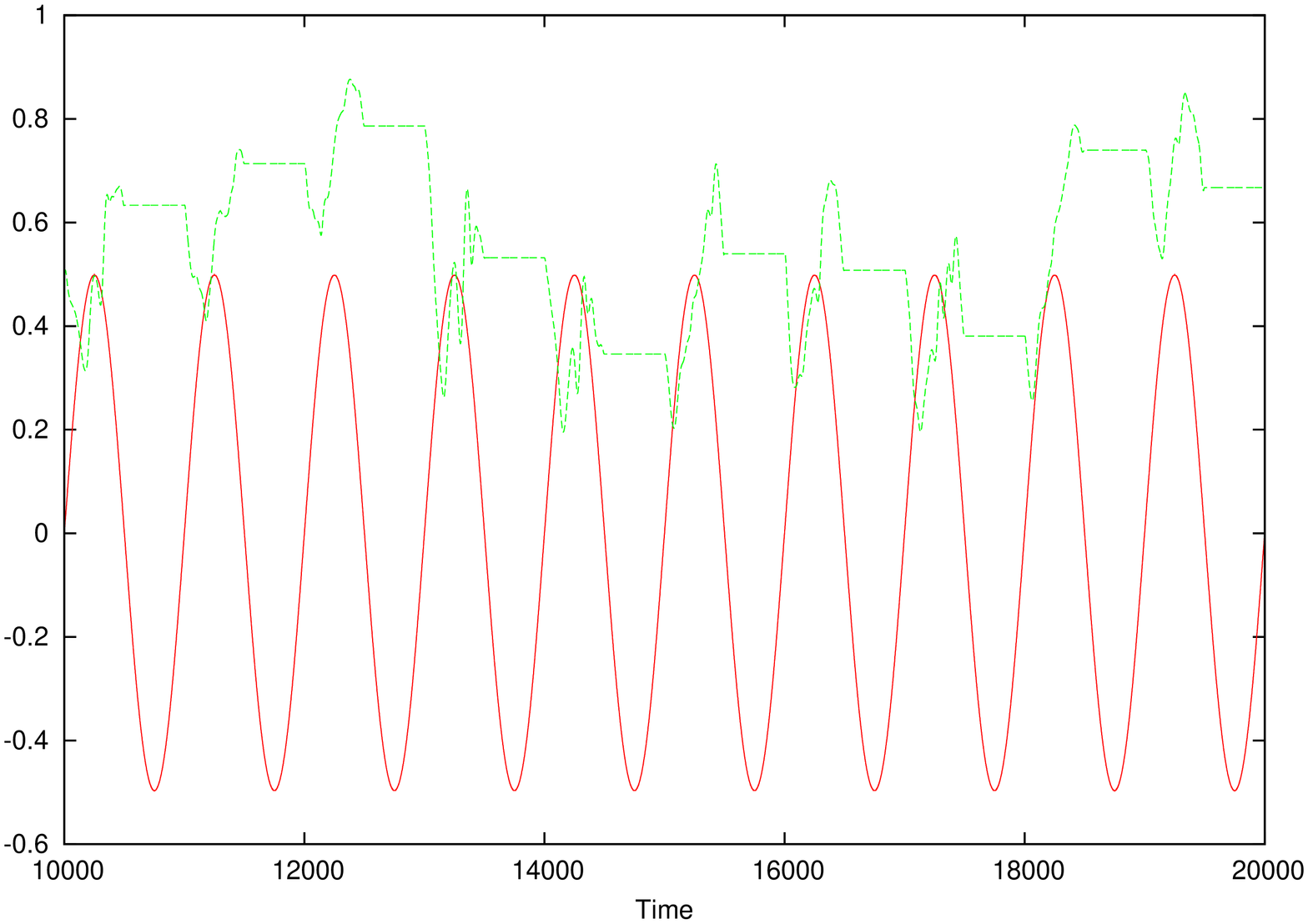}}

                  \end{tabular}
 \caption{\tiny{Phase Dynamics of the Oscillators with variable coupling radius (top) and the coevolution of the coupling radius $r$ (sinusoidal line) with order parameter $|Z|$ (non-sinusoidal line) (bottom). In both, the initial phases were random, $N=1000,\epsilon =0.07,\alpha =\frac{\pi }{2}-0.1$ and the coupling radius varied from nearest neighbor to global coupling with a time-period of $1000$ steps.}}
\end{center}
\end{figure}
This
leads to a hysteretic behaviour of $Z$ with $r$ also reported in one earlier work \cite{so1}, which can be related to an asymmetry in the formation and disruption of synchronization in the system. The hysteresis arises due to a competition between the intrinsic time scale of
the coupled phase oscillators and the time scales (time period of $R(t)$) of the time dependent couplings. The area of the hysteresis loop is found to increase with the coupling strength $\epsilon_0$ as is seen from the fig. $(7)$. However, all these behaviors are observed when the coupling strength is in the limit $\epsilon <0.1$. For higher coupling strength values, the system becomes globally synchronized in a short time time and stays there for ever. It is to be noted here that similar hysteretic behaviour is observed for time dependent strength of coupling also and is well-studied \cite{so1}.  
\begin{figure}[h]
\begin{center}
\begin{tabular}{c}
        \resizebox{6cm}{!}{\includegraphics[angle=-90]{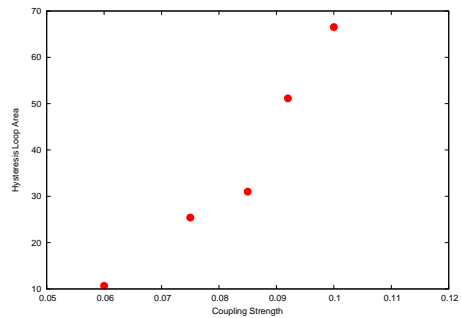}}
                  \end{tabular}
 \caption{\tiny{Variation of the Hysteresis Loop Area with the Coupling Strength ($\epsilon$) for $N=100,R_{0}=25,T_{r}=1000$.}}
\end{center}
\end{figure}

\vskip 1 cm

\noindent {\bf IV. SIMULATION OF THE LOW DIMENSIONAL SPATIO-TEMPORAL DYNAMICS USING OTT-ANTONSEN ANSATZ}
\vskip 0.5 cm

We now try to give an analytical justification of the hysteretic behavior of the order parameter presented in the previous section using the recently developed Ott-Antonsen ansatz \cite{ott1}. To use the ansatz, we go to the thermodynamic ($N\rightarrow \infty$) limit and consider the continuum limit description of space and time so that the Kuramoto equation (eqn. $1$) now becomes
 
\begin{equation}
\frac{\partial \theta(x,t)}{\partial t}=\omega-\frac{\epsilon}{2R}\int_{x-R}^{x+R}{\rm sin}(\theta_i(x,t)-\theta_j(x',t)+\alpha)dx'.
\end{equation}

To study the collective dynamics, we define the \textit{local} time-dependent complex order parameter $Z(x,t)$ by the following equation

\begin{equation}
Z(x,t)=\frac{\epsilon}{2R}\int_{x-R}^{x+R}e^{i\theta(x')}dx'.
\end{equation}

Next we define a distribution function for the oscillators $f(x,\omega,\theta,t)$ such that it gives the number of oscillators situated in the region between $x$ and $x+dx$, with their phases in the range $\theta$ to $\theta+d\theta$, having intrinsic angular frequency between $\omega$ to $\omega+d\omega$ at a time $t$. With this definition, we can rewrite eqn. $8$ as

\begin{equation}
Z(x,t)=\frac{\epsilon}{2R}\int_{x-R}^{x+R}\int_{-\infty}^{+\infty}\int_{-\pi }^{+\pi }e^{i\theta'}f(x',\omega, \theta ',t)d\theta'd\omega dx'.
\end{equation}

Due to the local conservation law of oscillator numbers, the distribution function is needed to satisfy the continuity equation,

\begin{equation}
 \frac{\partial f}{\partial t}=\frac{\partial}{\partial \theta}(fv)
\end{equation}
where the phase velocity $v$ is given by the relation
\begin{equation}
v=\omega -\frac{\epsilon}{2R}\int_{x-R}^{x+R}\int_{-\infty}^{+\infty}\int_{-\pi }^{+\pi }{\rm sin}(\theta_i(x)-\theta_j(x')+\alpha)d\Omega, 
\end{equation}
where $d\Omega =f(x',\omega,\theta ',t)d\theta'd\omega dx'$.
Using eqn. $(10)$, this can be written in a more compact form as

\begin{equation}
v=\omega -\frac{1}{2}[Ze^{-i(\theta +\alpha -\frac{\pi}{2})}+Z^{*}e^{i(\theta +\alpha -\frac{\pi}{2})}].
\end{equation}

Finally, we use the Ott-Antonsen ansatz \cite{ott1} to extract the dynamics in terms of only two equations. We note that due to the the $2\pi-$periodicity in the definition of the phases $\theta $, the distribution function of the phases $f$ can be written in the form of a usual Fourier series, with the Fourier components given by various powers of a function $a(x,\omega ,t)$ according to the ansatz, i.e.,

\begin{equation}
f(x,\omega,\theta,t)=\frac{g(\omega)}{2\pi}{1+\sum_{n=1}^{\infty}[a^{n}e^{in\theta}+(a^{*})^{n}e^{-in\theta}]}.
\end{equation}

Finally, substitution of eqns $(13)$ and $(14)$ to the eqns $(10)$ and $(11)$ gives along with our initial assumption of identical oscillators (i.e., $g(\omega )=\delta(\omega -\omega _{0})$ where the common intrinsic angular frequency $\omega _{0}$ was set to $0$ without loss of generality) the two coupled equations for determining the spatio-temporal dynamics of the local order parameter which are 

\begin{equation}
Z(x,t)=\frac{\epsilon}{2R}\int_{x-R}^{x+R}a^{*}(x',t)dx',
\end{equation}

and

\begin{equation}
\frac{\partial a}{\partial t}=\frac{1}{2}[Z^{*}e^{i\alpha }-a^{2}Ze^{-i\alpha }]
\end{equation}

where the $a$ now denote $a=a(x,t)=a(x,0,t)$. Finally, for our problem the $R$ varies sinusoidally in time and smoothly changes between $0$ and a given maximum value.

Thus we set the eqns $(15)$ and $(16)$ which are suitable for predicting the low-dimensional behavior of the Kuramoto system. In this regard, we mention that the spatio-temporal dynamics of the Ott-Antonsen equations for fixed coupling has been analytically studied only very recently \cite{wolf1} and is reported to demonstrate interesting dynamical features like plane wave solutions, modulational instabilities, amplitude and phase turbulence. However, such a study for time-dependent coupling is still lacking.  
\begin{figure}[h]
\begin{center}
\begin{tabular}{c}
         \resizebox{!}{!}{\includegraphics[width=4cm,height=5cm, keepaspectratio,angle=-90]{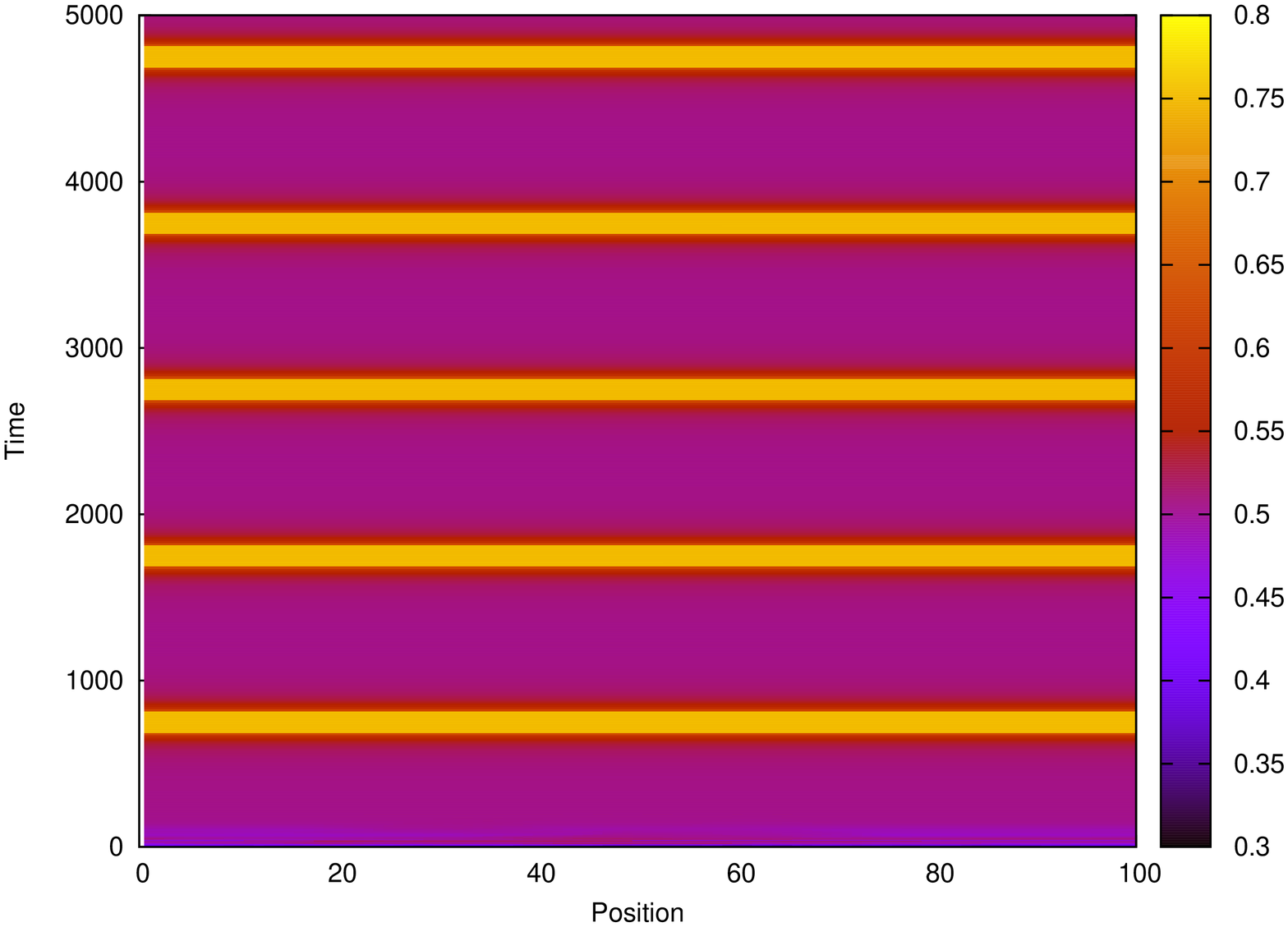}}\\
        \resizebox{!}{!}{\includegraphics[width=4cm,height=5cm, keepaspectratio,angle=-90]{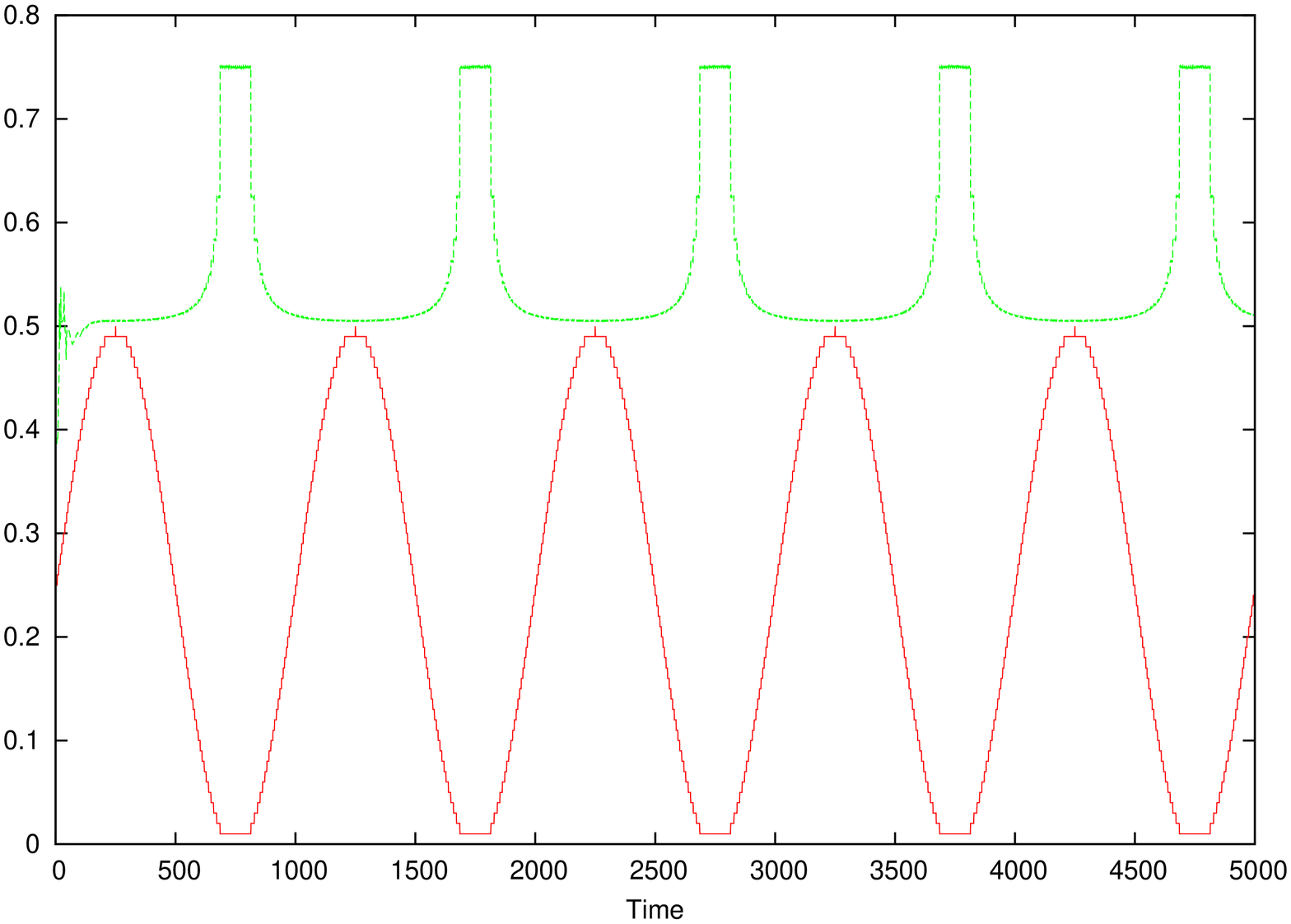}}
        
                  \end{tabular}
 \caption{\tiny{Time Evolution of the Modulus of the Local Order Parameter $|Z|$ in Space and Time (top) and Coevolution of the Coupling Radius $r$ (sinusoidal line) and $|Z(20,t)|$ (non-sinusoidal line) in Time (bottom) for $\alpha=\frac{\pi}{2}-0.1, \epsilon=0.5$, the couplings oscillating between nearest neighbour to global coupling within a time period of $1000$ unit.}}
\end{center}
\end{figure}

\begin{figure}[h]
\begin{center}
\begin{tabular}{c}
         \resizebox{!}{!}{\includegraphics[width=4cm,height=5cm, keepaspectratio,angle=-90]{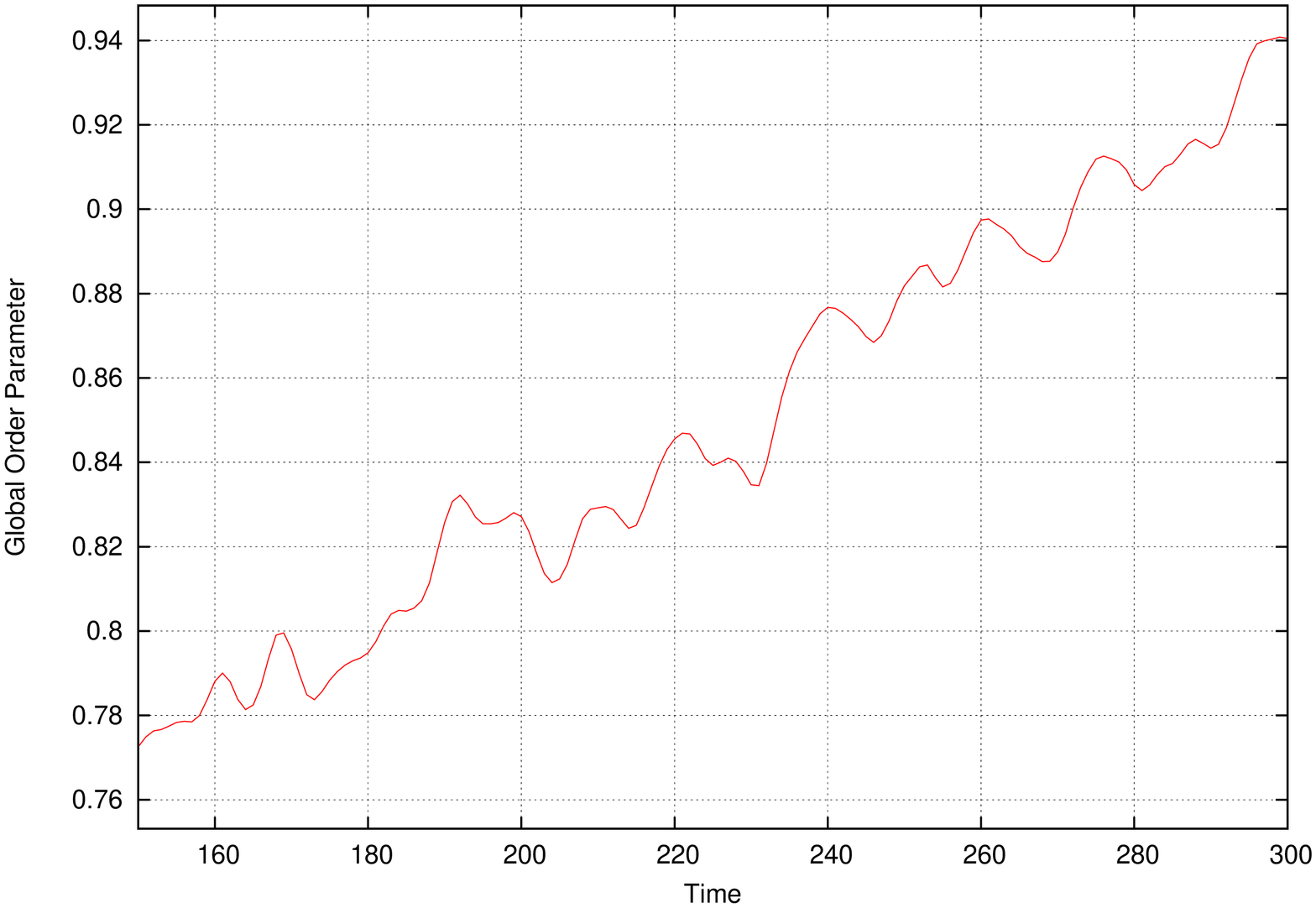}}
        
                  \end{tabular}
 \caption{\tiny{Time-Evolution of the Global Order Parameter Obtained from Direct Simulation of Kuramoto model with $\alpha=\frac{\pi}{2}-0.1, \epsilon=0.5, L=1000$ and $dt=1.0$. Here the expected Intrinsic Time Scale is $\frac{2\pi }{\epsilon sin\alpha }\approx 13$ and the observed oscillation is also seen to occur with a period roughly equal to that time scale.}}
\end{center}
\end{figure}

We proceed to solve eqns $(15)$ and $(16)$ numerically using discretization over space with $dx=1$ (such that the integral in eqn. $(15)$ becomes a sum over discrete values) and time with $dt=0.01$ and use Euler's method of iteration to eqn. $(16)$. As the initial conditions, the initial order parameter ideally should be exactly zero everywhere as in our simulations described in the previous sections, the initial phases were independently and randomly selected from the range $[0,2\pi]$. However, $Z=0,a=0$ is a stationary state for the eqns. $(15)$ and $(16)$. So to observe the dynamics we start with some finite values for $a(x,0)$ choose real and imaginary parts of $a(x,0)$ randomly from the range $[0,1]$ independently for each $x$ value. The numerical solutions show that, for a slowly changing coupling radius, we have a a spatially homogeneous local order parameter which oscillates in time with an amplitude which increases with the coupling strength $\epsilon$ (this is shown in the fig. $(8)$). 
This indicates that the system itself remains homogeneous at all times. This is expected -- as the coupling radius changes slowly, so at each step of its change the system gets sufficient time to homogenize and reach equilibrium at the instantaneous coupling radius value. 
Indeed the system of eqns $(15)$ and $(16)$ admits spatially homogeneous but time-dependent oscillatory solution given by 

\begin{equation}
a(t)=e^{i(\epsilon sin\alpha)t}
\end{equation}

and

\begin{equation}
Z(t)=\epsilon e^{-i(\epsilon sin\alpha)t}
\end{equation}
 which indicates that the system has an intrinsic time scale of the order of $\frac{2\pi }{\epsilon sin\alpha }$. We also confirm the existence of this time-scale in the system by direct simulation of the coupled equations $(2)$ and tracking the time-evolution of the global order parameter which shows an oscillatory behavior with that time-scale (fig. $(9)$). Now if the time scale associated with the switch in coupling is much larger than this, then the system is able to follow the variable coupling while still maintaining its spatial homogeneity.
\begin{figure}[h]
\begin{center}
\begin{tabular}{c}
        \resizebox{5cm}{!}{\includegraphics[width=5cm,height=5cm,angle=-90]{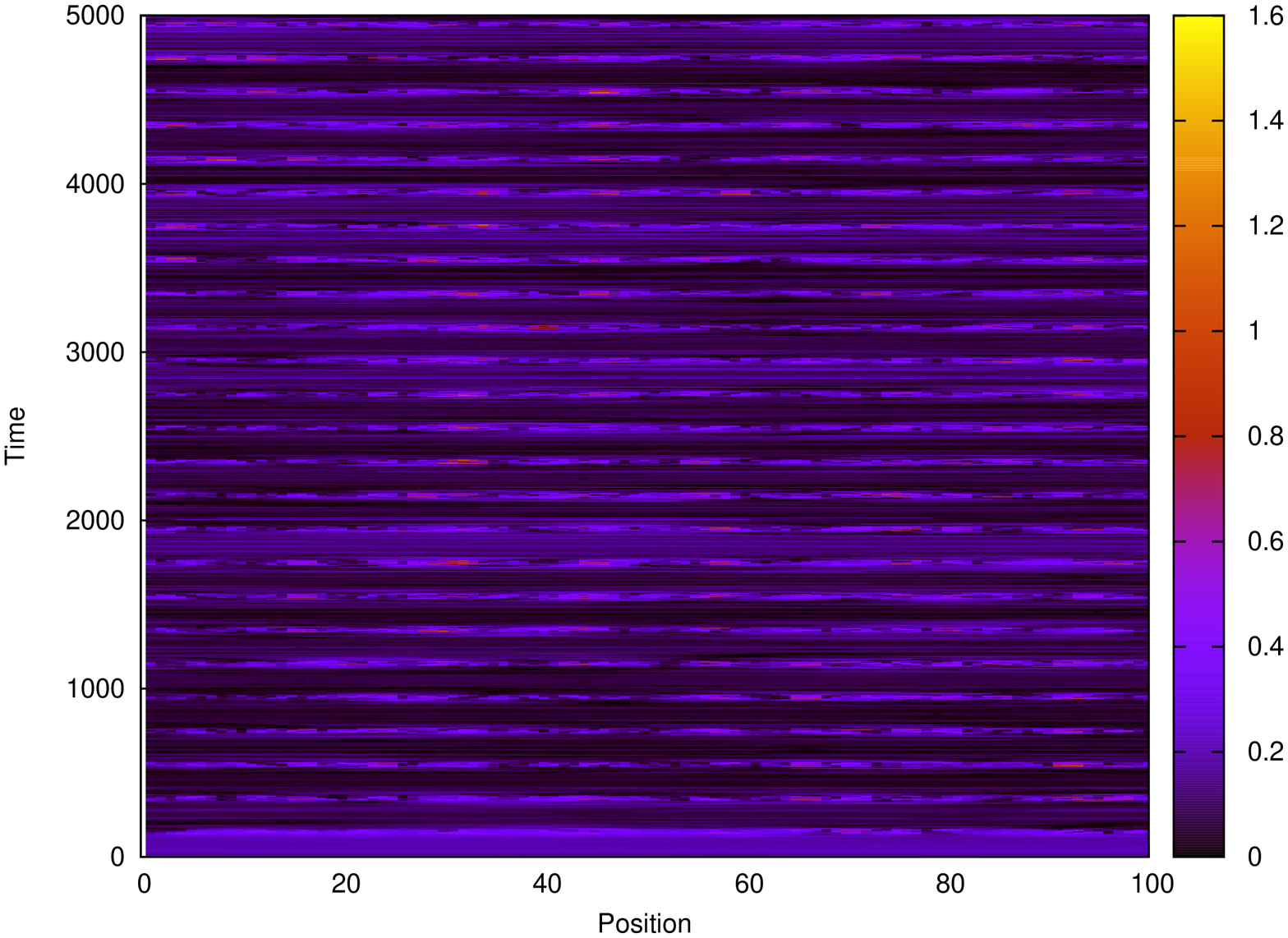}}\\
        \resizebox{4.5cm}{!}{\includegraphics[angle=-90]{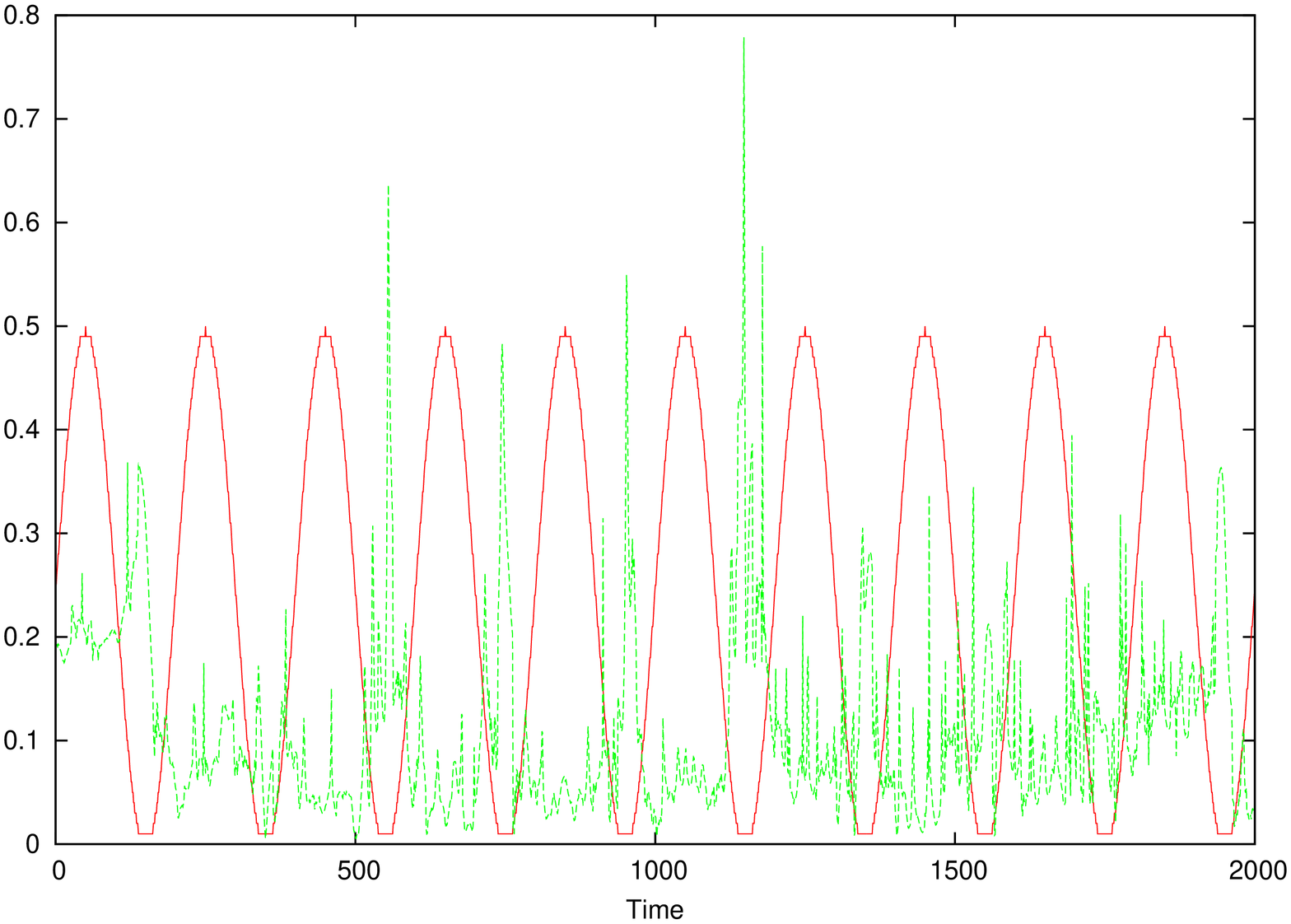}}
        \resizebox{4.2cm}{!}{\includegraphics[angle=-90]{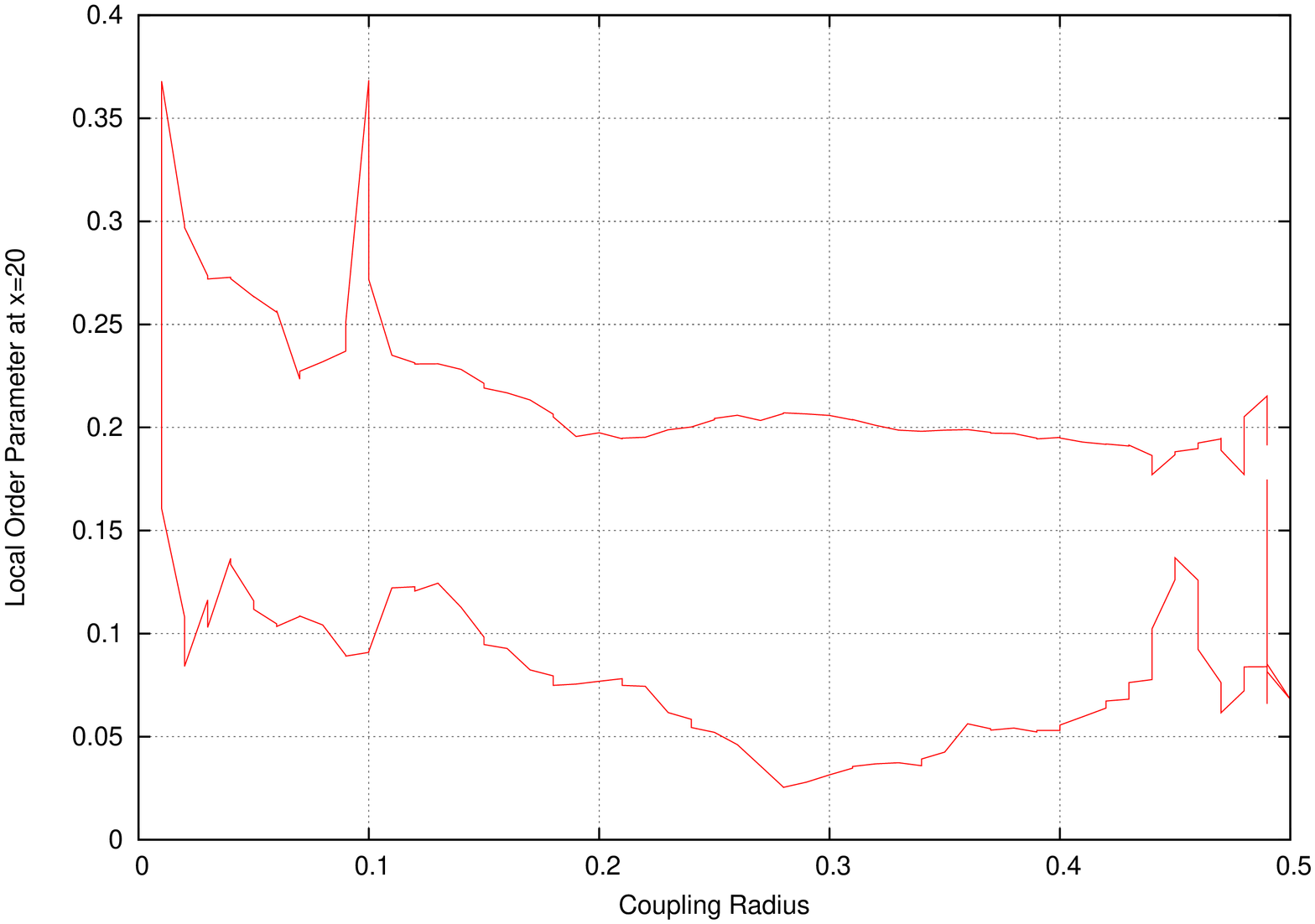}}
        
                  \end{tabular}
 \caption{\tiny{Time Evolution of the Modulus of the Local Order Parameter $|Z|$ in Space and Time (top) and Coevolution of the Coupling Radius $r$ (sinusoidal line) and $|Z(20,t)|$ (non-sinusoidal line) in Time (bottom left) for $\alpha=\frac{\pi}{2}-0.01, \epsilon=0.25$, and the associated hysteresis loop (bottom right) the couplings oscillating between nearest neighbour to global coupling within a time period of $200$ unit.}}
\end{center}
\end{figure}
  The rise and fall of the local order parameter are almost perfectly symmetric in each period and very regular over time and so no hysteresis is observed in this case. Also, in this limit, in the analysis of the system, one can replace time-varying coupling with a fixed coupling strength effectively by a uniform global coupling but with the coupling strength varying with time, a situation earlier studied \cite{spa1}. In those cases, however, it is possible to observe hysteretic phenomena if one considers, for example, a Lorentzian distribution for the intrinsic angular frequencies of the oscillators (as considered by Ott and Antonsen in their original work \cite{ott1}). The Ott-Antonsen equations in that case reduce to the following equation for the amplitude $\rho $ of the order parameter \cite{ott1} 
  
 \begin{equation}
  \frac{d\rho}{dt}=\left(1-\frac{\epsilon }{2}\right)\rho + \frac{1}{2}K\rho ^{3}
 \end{equation}
 which is nothing but the well-known mean-field dynamics for the magnetization $m$ of an Ising magnetic sample governed by the equation
 
 \begin{equation}
\frac{dm}{dt}=-\frac{\partial f}{\partial m} 
 \end{equation}
with the free energy $f(m)$ given by the Landau phenomenological equation
 
\begin{equation}
 f(m)=\left(1-\frac{\epsilon }{2}\right)m^{2} + \frac{1}{2}Km ^{4}.
\end{equation} 
  In this case, $\epsilon $ is identifiable with the temperature of the sample (indeed, the boundary $\epsilon =2$ between synchronous and the asynchronous state is analogous the critical temperature $T_{c}$ for the ferro-para transition). If this parameter is varied periodically with time, one can observe a hysteretic dynamics of $\rho $ with $\epsilon $ which is analogous to the thermal hysteresis of a ferromagnetic sample.
 
 However, returning to our present study, we note that when the frequency of the variable connectivity is of the order of $\epsilon sin\alpha$, the system fails to follow the changing connectivity. Its dynamics then become much faster than that in the previous case and as a result, spatial homogeneity is lost (fig. $(10)$). The order parameter now varies from point to point within the system. Unlike the previous limit, the rise and fall of the order parameter becomes extremely asymmetric and irregular in time and hence a hysteretic behavior with the coupling radius is observed. A full analytical treatment of the transition leading to the instability of the spatially homogeneous solution is worth pursuing.

\vskip 1 cm
 
\noindent {\bf V. SUMMARY AND CONCLUSION}
\vskip 0.5 cm
In this work, we have obtained a detailed analysis of the Kuramoto model with various kinds of time-dependent coupling range. To start with, we have characterized the entire parameter space of the model using the statistical measures of incoherence strength and discontinuity. At first, we have considered the phase dynamics of the oscillators when there is only a single switching event for the coupling radius with the condition that when the switching occurs, the system is in a `twisted' state with a certain periodicity. It is seen that the final state that emerges, has interesting spatial correlation property of the phases which depend on the initial coupling radius. We also studied the Kuramoto dynamics under sinusoidally changing coupling and demonstrated the existence of hysteretic behavior of the order parameter with the coupling range with the area of the hysteresis loop generally diminishing with decreasing coupling strength. This hysteresis points to the existence of an intrinsic time-scale of the system itself, which is confirmed both by the direct simulation of the coupled system $(2)$ and an analytic exact solution of the Ott-Antonsen ansatz that is spatially homogeneous. Indeed, the simulation of the Ott-Antonsen equations show that when there is a competition between this intrinsic time-scale to the time-scale of the switching coupling, the system no longer stays spatially homogeneous.
We shall carry out another numerical study in the future with fourth-order Runge-Kutta method and shall find out the magnitude of the time-scale more accurately.
 
This work with variable coupling can be further continued to other nonlinear coupled oscillators like Lorenz, Stuart-Landau or Rossler oscillators and can be tried in systems with various coupling topology for interesting results.

------------------------------------------------------------

\end{document}